\documentclass[aps,showpacs,amssymb,amsmath,amsfonts,pra,12pt]{revtex4-1}
\pdfoutput=1
\usepackage{amsfonts}
\usepackage{amssymb}
\usepackage{amsmath}
\usepackage{graphicx}
\usepackage{bbm}
\usepackage{overpic}
\usepackage{subfigure}
\usepackage{xcolor}

\newtheorem{lemma}{Lemma}

\newtheorem{proposition}{Proposition}

\newcommand{\proof}{\noindent {\it Proof.\ }}
\newcommand{\scalar}[2]{\langle #1 | #2 \rangle}
\newcommand{\ketbra}[2]{| #1 \rangle \langle #2 |}
\newcommand{\ket}[1]{| #1 \rangle}
\newcommand{\bra}[1]{\langle #1 |}
\newcommand{\proj}[1]{| #1 \rangle \langle #1 |}
\newcommand{\kron}{\otimes}
\newcommand{\tr}{\mathrm{tr}}
\newcommand{\1}{{\rm 1\hspace{-0.9mm}l}}
\newcommand{\Id}{\1}
\newcommand{\halmos}{\hfill $\blacksquare$\newline}
\newcommand{\R}{\ensuremath{\mathbb{R}}}

\newcommand{\setN}{\ensuremath{\mathbb{N}}}
\newcommand{\diag}{\mathrm{diag}}

\newcommand{\Cplx}{\mathbb{C}}

\newcommand{\vE}[1]{\ensuremath{S\left(#1\right)}}
\newcommand{\MOE}[1]{\ensuremath{S_{\min}\left(#1\right)}}
\newcommand{\States}[1]{\ensuremath{\Omega_{#1}}}

\newcommand{\LambdaRm}{\mathrm{\Lambda}}
\newcommand{\NumRange}[1]{\ensuremath{\LambdaRm({#1})}}
\newcommand{\CNumRange}[1]{\ensuremath{\LambdaRm^C({#1})}}
\newcommand{\ProductNumRange}[1]{\ensuremath{\LambdaRm^{\!\otimes}\! \left( #1 \right)}}

\newcommand{\lambdaProdMin}{\lambda_{\rm min}^{\otimes}}
\newcommand{\lambdaProdMax}{\lambda_{\rm max}^{\otimes}}

\newcommand{\eg}{\emph{e.g.}}
\newcommand{\cf}{\emph{cf.}}
\newcommand{\etal}{\emph{et al.}}

\newcommand{\mprod}{\boxtimes}

\renewcommand{\bar}[1]{{#1}^*}

\usepackage{xcolor}

\begin{document}

\title{Restricted numerical range: 
            a versatile tool in the theory of quantum information 
}
\author{Piotr Gawron}

\author{Zbigniew Pucha{\l}a}

\author{Jaros{\l}aw Adam Miszczak}
\affiliation{Institute of Theoretical and Applied Informatics, Polish Academy
of Sciences, Ba{\l}tycka 5, 44-100 Gliwice, Poland}

\author{{\L}ukasz~Skowronek}
\affiliation{Instytut Fizyki im. Smoluchowskiego, Uniwersytet
Jagiello{\'n}ski, Reymonta 4, 30-059 Krak{\'o}w, Poland 
}
\author{Karol~{\.Z}yczkowski}
\affiliation{Instytut Fizyki im. Smoluchowskiego, Uniwersytet
Jagiello{\'n}ski, Reymonta 4, 30-059 Krak{\'o}w, Poland 
}
\affiliation{Centrum Fizyki Teoretycznej, Polska Akademia Nauk, Aleja Lotnik{\'o}w 
32/44, 02-668 War\-sza\-wa, Poland
}

\date{August 14, 2010}

\begin{abstract}
Numerical range of a Hermitian operator $X$ is defined as the set of all
possible expectation values of this observable among a normalized quantum state.
We analyze a modification of this definition in which the expectation value is
taken among a certain subset of the set of all quantum states. One considers for
instance the set of real states, 
the set of product states, separable states, or the set of maximally entangled
states. We show exemplary applications of these algebraic tools in the theory of
quantum information: analysis of $k$--positive maps and entanglement witnesses,
as well as study of the minimal output entropy of a quantum channel. Product
numerical range of a unitary operator is used to solve the problem of local
distinguishability of a family of two unitary gates.
\end{abstract}
\pacs{03.65.-w, 03.67.-a, 02.10.Yn}

\maketitle

\section{Introduction}

Expectation value of a Hermitian observable $X$ among a given 
pure state $|\psi\rangle$ belongs to the basic notions of quantum theory. It is
easy to see that the set $\Lambda$ of all possible expectation values of a given
operator $X$ among all normalized states forms a close interval between the
smallest and the largest eigenvalue,
$\Lambda(X)=[\lambda_{\rm min}, \lambda_{\rm max}]$.

In the theory of matrices and operators one calls such a set
{\it numerical range} or {\it field of values} of an operator $X$,
which in general needs not to be Hermitian \cite{HJ2,gustav}.
Properties of numerical range are intensively studied in the mathematical 
literature \cite{ando94numerical,li95cnumerical},
several generalizations of this notion were investigated
\cite{Ma73,MW80,BLP91,LZ01},
and its usefulness in quantum theory has been emphasized \cite{KPLRS09}.

Let us introduce the set $\Omega$ of all density matrices of size $N$, which are
Hermitian, positive and normalized, $\Omega:=\{\rho: \rho^{\dagger}=\rho\ge 0,
\quad {\rm Tr}\rho=1\}$. If a given state is pure, $\rho=|\psi\rangle \langle
\psi|$, the expectation value reads Tr$\rho X=\langle \psi |X|\psi\rangle$. Any
density matrix can be represented as a convex combination of pure states. Hence
for any operator the sets of its expectation values among pure states and among
mixed states are equal.

More formally, let $X$ be an arbitrary operator acting on an $N$-dimensional 
Hilbert space ${\cal H}_N$. Its \emph{numerical range},
can be defined as
\begin{equation}
\Lambda(X)=\{ {\rm Tr} X\rho, \ \rho\in \Omega \}.
\label{range1}
\end{equation}
The related concept of \emph{numerical radius}
\begin{equation}\label{numraddef}
 r\left(X\right)=\left\{\left|z\right|,\ z\in\Lambda\left(X\right)\right\}
\end{equation}
is also a frequent subject of study \cite{ando94numerical,li95cnumerical} (\cf\ Table \ref{tab2} in Section \ref{sec:concl}).

In this paper we analyze a modification of the standard definitions \eqref{range1} and \eqref{numraddef}. For any
operator $X$, one defines its {\it restricted numerical range}
\begin{equation}
\Lambda_R(X)=\{ {\rm Tr} X\rho, \ \rho\in \Omega_R \subset \Omega \}.
\label{rest1}
\end{equation}
and the {\it restricted numerical radius}
\begin{equation}\label{defrestnumrad}
r_R\left(X\right)=\left\{\left|z\right|,\ z\in\Lambda_R\left(X\right)\right\}.
\end{equation}
The symbol $\Omega_R$ denotes an arbitrary subset of the set $\Omega$ of all
normalized density matrices of size $N$.
Thus the above definition of the restricted
numerical range is more general than the one
studied in \cite{DMS87,gustav}, 
in which a subset of the set of pure states was used.

Some examples of restricted numerical ranges are listed in Tab. \ref{tab1}. The
range restricted to real states was recently discussed by Holbrook \cite{Ho09},
while the Liouville numerical range, in which the pure states of size $M^2$
reshaped into a square matrix form a legitimate density operator was analyzed by
Silva \cite{Si09}. The numerical range of a density matrix $\rho$ restricted to
the $SU(2)$ coherent states gives the set of values taken by its Husimi
representation - see e.g. \cite{BZ06}. Examples of restricted numerical radii can
be found in Tab. \ref{tab2} at the end of the paper. An very important example is the product numerical radius $r^{\otimes}(X) = \max\{|\bra{\psi_1\otimes\ldots\otimes\psi_m}X
\ket{\psi_1\otimes\ldots\otimes\psi_m}|:
\ket{\psi_i}\in\mathcal{H}_{n_i}\}$, which coincides for $m=2$ and $X$ normal with the
Schmidt operator norm $\left\|X\right\|_{S(1)}$ introduced by Johnston and Kribs \cite{JK09}.

If the dimension of the Hilbert space is a composite number, $N=KM$, the space
can be endowed with a tensor product structure
\begin{equation}
\mathcal{H}_N = \mathcal{H}_K \otimes \mathcal{H}_M.
\label{HKM}
\end{equation}
From a physical perspective this corresponds to distinguishing two
subsystems in the entire system.
One defines then the set of separable pure states, i.e. the states
with the product structure, 
$|\psi\rangle=\ket{\psi_A}\otimes \ket{\psi_B}$.

Substituting this set into definition 
(\ref{rest1}) of the restriced product range
one arrives at the notion of {\it product numerical range}
of an operator $X$,
\begin{equation}
  \ProductNumRange{X} = \left\{ \bra{\psi_A \otimes \psi_B} X
   \ket{\psi_A\otimes\psi_B} 
: \ket{\psi_A}\in \mathcal{H}_K, \ket{\psi_B}\in \mathcal{H}_M \right\},
   \label{lnr}
\end{equation}
where both states 
$\ket{\psi_A} \in {\cal H}_K$  and $\ket{\psi_B} \in {\cal H}_M$ are normalized.

The product numerical range can also be
considered as a particular case of the
 decomposable numerical range \cite{Ma73,MW80}
defined for operators acting on a tensor product Hilbert space.
This notion was recently analyzed in 
\cite{dirr08relative,DFY08,thomas08significance,SHGDH08},
where the name  \emph{local numerical range} was used.
In physics context the word `local' refers to local action, 
so the unitary matrix with a tensor
product structure, $U(M)\otimes U(K)$, is said
to act `locally' on both subsystems.
To be consistent with the mathematical terminology we will use here
the name ``product numerical range'', 
although a~longer version  ``local product numerical range'' 
would be even more accurate.
Note that one may also use other restricted sets of quantum states
as these mentioned in Table \ref{tab1}.

\begin{table}\label{tbl:def1}
\scalebox{0.95}{\begin{tabular}{|l|l|c|}
\hline
\hline
Restricted NR  & $\Omega_R \subset \Omega
:=\{\rho: \rho^{\dagger}=\rho\ge 0, \, {\rm Tr}\rho=1\}$.
 & dimension $N$ \\
 \hline \hline 
 NR restricted to real states &  
$\Omega_R=\{|\psi\rangle \langle \psi|, \, |\psi\rangle \in {\mathbbm R}^N\}$ & arbitrary  \\ \hline
 Product NR &  
$\Omega_R=\{|\psi\rangle \langle \psi|,\,  |\psi\rangle =|\phi^A \otimes \phi^B \rangle \} $ &    $K \times M$ \\ \hline
 Separable NR &  
$\Omega_R=\{\sum_i p_i | \psi_i\rangle \langle \psi_i|,\, 
 |\psi_i\rangle = |\phi^A_i \otimes \phi^B_i \rangle\}$ &  
  $K \times M$ \\ \hline
 Schmidt Rank $k$ NR &  
$\Omega_R=\{\sum_i p_i |\psi_i\rangle \langle \psi_i|,\, 
 |\psi_i\rangle = \sum_{j=1}^k q_{ij} |\phi^A_{ij} \otimes \phi^B_{ij} \rangle\}$ &  
  $K \times M$ \\ \hline
 Liouville NR &
$\Omega_R=\{|\psi\rangle \langle \psi|, \,  |\psi\rangle =\sum_{ij}\sigma_{ij}|i,j\rangle, \, 
\sigma^{\dagger}=\sigma\ge 0, {\rm Tr}\sigma=1\}$ &
 $M \times M$ 
\\ \hline
 $SU(K)$ coherent states NR &  
$\Omega_R=\{|\psi\rangle \langle \psi|, \, |\psi\rangle \in SU(K) 
{\rm \ coherent \ states} \}$
& $\frac{(K+l-1)!}{l! (K-1)!}, \ l\in\setN$
\\ \hline  \hline 
\end{tabular}}
\caption{
Examples of {\emph {restricted}} numerical range (NR):
 $\Lambda_R(X)=\{ {\rm Tr} X\rho, \rho\in \Omega_R  \}$, 
where $\Omega_R \subset \Omega$ denotes a subset of the set of all
quantum states of size $N$. All pure states are assumed to be normalized,
$\langle \psi|\psi\rangle=1$, while all coefficients in the sums are non-negative. In each case eq. \eqref{defrestnumrad} provides example of the corresponding notion of restricted numerical radius $r_R$.
}
\label{tab1}
\end{table}

The main aim of this work is to demonstrate usefulness of the restricted
numerical range for various problems of the theory of quantum information. This
paper is organized as follows. In Section \ref{sec:product1} we review some
basic features of product numerical range and present some examples obtained for
Hermitian and non-Hermitian operators.
Although we mostly discuss the simplest case of a two-fold tensor product
structure, which describes the physical case of a bi-partite system, we analyze
also operators representing the multi-partite systems. In Section
\ref{sec:separable} we study the notion of {\it separable numerical range} and
other {\it restricted} numerical ranges of an operator acting on a composed
Hilbert space.

Key results of this work are presented in Sec. \ref{sec:qinfo} in which some
applications in the theory of quantum information are presented. In particular,
by analyzing a family of one-qubit maps we find the conditions under which the
map is positive and establish a link between product numerical range of a
Hermitian operator and the minimum output entropy of a quantum channel. The
problem of $k$--positivity of a quantum map is shown to be connected with
properties of the numerical range of the corresponding Choi matrix restricted to
the set $\Omega^{({\rm k})}$ of states with the Schmidt number not larger than
$k$ \cite{TH00}. For $k=2$, we point out that the question of distillability of
an entangled quantum state is related to the numerical range restricted to the
set $\Omega^{({2})}$.

Furthermore, properties of product numerical range of non-Hermitian operators
are used to solve the problem of local distinguishability for a family of
two-qubit gates. In section \ref{sec:concl} we present some concluding remarks
and discuss further possibilities of generalizations of numerical range which
could be useful in quantum theory. Proofs of certain lemmas are relegated to the
Appendix.

\section{Product numerical range}
\label{sec:product1}

Quantum information theory deals with composite quantum systems which can be
described in a complex Hilbert space with a tensor product structure
\cite{Kr05}. When analyzing properties of operators acting on the composed
Hilbert space (\ref{HKM}), it is physically justified to distinguish
\emph{product} properties, which reflect the structure of the Hilbert space.

If the physical system is isolated from the environment, its dynamics in time
can be described by a unitary evolution $\ket{\psi'} = U\ket{\psi}$ where $U$ is
unitary, $UU^{\dagger}=\1_N$. In the case of a bipartite system, $N=KM$, one
distinguishes a class of \emph{local dynamics}, which take place independently
in both physical subsystems, so that $U=U_A \otimes U_B$, where $U_A \in U(K)$,
while $U_B \in U(M)$. From a group-theoretical perspective, one distinguishes
the direct product $U(K)\times U(M)$, which forms a proper subgroup of $U(KM)$.

It is important to know which tasks, such as the discrimination of pure quantum
states, can be completed with the use of local operations and classical
communication. For this purpose, it is convenient to work with the notion of the
product numerical range of an operator defined by Eq. (\ref{lnr}). This
algebraic tool can be considered as a natural generalization of the standard
numerical range for operators acting on a tensor product Hilbert space.

Note that the definition of \emph{product numerical range} is not unitarily
invariant, but implicitly depends on the particular decomposition of the Hilbert
space. This notion may also be considered as a~numerical range \emph{relative}
to the proper subgroup $U(K)\times U(M)$ of the full unitary group $U(KM)$. It
is worth mentioning that product numerical range differs from so-called
\emph{quadratic numerical range}, also defined for operators acting on a
composite Hilbert space \cite{LMMT01}.

\medskip

Consider the following problems arising in the theory of quantum information.
\begin{itemize}
\item[$\bullet$]
 Verify if a given map $\Phi$ acting on the set of quantum states is positive:
 Is $\Phi(\rho)\ge 0$  for all $\rho\ge 0$?
\item[$\bullet$]
 For a given observable $X$, defined for a bipartite system,
 find the largest (the smallest) expectation value among pure product states: 
 What  is  {\rm max}$_{|\phi_A,\phi_B\rangle}$
 $\langle \phi_A,\phi_B|X|\phi_A,\phi_B\rangle$?
\item[$\bullet$]
 Check if two unitary gates $U_1$ and $U_2$ acting 
on a bipartite systems are distinguishable.
 This is the case if there exists a product 
  state  $| \psi\rangle=|\phi_A,\phi_B\rangle$ such that
 the states $U_1|\psi\rangle$ and $U_2|\psi\rangle$ are orthogonal.
\item[$\bullet$]
 For a pair of two bipartite states $\sigma$
 and $\rho$  maximize their fidelity or the trace Tr$\rho\sigma$
by the means of local operations.
\end{itemize}

This list of questions, of different difficulty levels, could be easily
extended. All these problems have one thing in common: they could be directly
solved, if we had an efficient algorithm to compute the product numerical range
of an operator. Although in this work we are not in a
position to go so far, we aim to show usefulness of this notion
and present some partial results.



\subsection{Basic properties}
\label{sec:basic}
In this section we review some basic properties of product numerical range.
Some of them were discussed by  Dirr \etal\ \cite{dirr08relative},
while some other were established in \cite{PGMSCZ10}.

For any operator $X$ acting on a Hilbert space ${\cal H}_N$,
its product numerical range (\ref{lnr}) forms a nonempty, connected set
in the complex plane.
However, this set needs not to be convex nor simply connected.
Further properties of product numerical range include
\medskip

a) Subadditivity, \quad
$ \ProductNumRange{A+B}\subset \ProductNumRange{A}+\ProductNumRange{B}$,

\smallskip
b) Translation: \quad for any $\alpha\in \Cplx$ one has
$\ProductNumRange{A+\alpha \Id}=\ProductNumRange{A}+\alpha$,

\smallskip
c) Scalar multiplication: \quad
	for any  $\alpha\in \Cplx$ one has
$	\ProductNumRange{\alpha A}=\alpha\ProductNumRange{A}$,

\smallskip
d) Product unitary invariance: \quad 
$ \ProductNumRange{(U\otimes V)A(U\otimes V)^\dagger}=\ProductNumRange{A}$,

\smallskip
e) If $A$ is normal, then numerical range of its tensor product with an arbitrary operator $B$ coincides with the convex hull of the product numerical range,
$\NumRange{A \kron B} = \mathrm{Co}(\ProductNumRange{A \kron B})$

\smallskip
f) Product numerical range of any $A$ contains the {\it barycenter}
 of the spectrum,   $\frac{1}{KM}\; {\tr} A \ \in \  \ProductNumRange{A}$.

\medskip
To analyze product numerical range of the Kronecker product 
it is convenient to make  use of the geometric algebra of complex sets \cite{FMR01}.
For any two  sets $Z_1$ and $Z_2$ on the complex plane, one defines their 
\emph{Minkowski product},
\begin{equation}
Z_1 \mprod  Z_2 =  
\left\{z: z=z_1 z_2, \ z_1 \in Z_1, \ z_2 \in Z_2 \right\} .
\label{Mprod} 
\end{equation}
Observe that this operation is not denoted by the standard symbol $\otimes$
in order to avoid the risk of confusion with the tensor product of operators.
The above definition allows us to express 
the product numerical range of the Kronecker product of 
arbitrary two operators as a Minkowski product 
of the numerical ranges of both factors \cite{gustav}, 
\begin{equation}
  \label{rangetensorpr}
\ProductNumRange{A \otimes B}   =  \NumRange{A} \mprod  \NumRange{B},
\end{equation}
This  property can be directly 
generalized to an arbitrary number of factors.
Thus the problem of finding the product numerical range of a tensor product 
can be reduced to finding the Minkowski product \cite{FMR01,FP02}
of two or more numerical ranges.

\subsection{Hermitian case}
\label{sec:lr-Hermitian}

In the case of a Hermitian operator  $X=X^{\dagger}$ acting on ${\cal H}_N$
its spectrum belongs to the real axis. Labeling the eigenvalues
in a weakly increasing order, $\lambda_1 \le \lambda_2 \le \dots \le \lambda_N$,
one can write the numerical range as an interval, 
$\NumRange{X}=[\lambda_1,\lambda_N]$, see e.g. ~\cite{HJ2}.

Let us assume, the Hilbert space has a product structure, 
$\mathcal{H}_N=\mathcal{H}_K \otimes \mathcal{H}_M$,
which implies a notion of a pure product state.
Define the points $\lambdaProdMin$ and
$\lambdaProdMax$ as the maximal and the minimal
expectation values of $X$ among all product pure states.
Then the product numerical range is given by a closed interval,
$\ProductNumRange{X}=[\lambdaProdMin,\lambdaProdMax]$.
If the spectrum of $X$ is not degenerated to a single point,
 (which is the case iff $X$ is proportional to identity),
 then $\lambdaProdMin \ne \lambdaProdMax $,
so the product numerical range has a non-zero volume \cite{PGMSCZ10}.

Making use of the lemma about the dimensionality  of subspaces belonging to 
a composed Hilbert space of size $N=KM$ which contain at least one separable state
\cite{CMW08}, one can  get the following bounds for the edges of the
product numerical range
\begin{equation}
\label{hermbounds}
\lambdaProdMin \le \lambda_{(K-1)(M-1)+1}
{\quad \rm and \quad}
\lambdaProdMax \ge \lambda_{K+M-1}.
\end{equation}
These bounds, proved in \cite{PGMSCZ10}, imply that
in the simplest case of a $2 \times 2$ system ($N=4$),
the product numerical range contains the central segment of the
spectrum,
\begin{equation}
\label{herm22}
\Lambda(X)=[\lambda_1,\lambda_4] \supset
\ProductNumRange{X}=[\lambdaProdMin,\lambdaProdMax]
\supset [\lambda_2,\lambda_{3}] .
\end{equation}
Similarly, for any Hermitian $X$ acting on a $2\times K$ space
the central segment of the spectrum
$[\lambda_K,\lambda_{K+1}]$ belongs to 
$\ProductNumRange{X}$.
In the case of a $3\times 3$ system ($N=9$), the
product numerical range  of $X$ contains its central eigenvalue, 
$\lambda_5 \in \ProductNumRange{X}$.

\subsubsection{Exemplary Hermitian matrix of order four}

Not being able to construct an algorithm 
to obtain product numerical range for an arbitrary Hermitian operator
we shall study some concrete examples.
Consider first positive numbers $t,s \geq 0$ 
and a family of Hermitian matrices of order four
\begin{equation}
\label{xts}
X_{t,s} = 
\left(
\begin{array}{cccc}
2 & 0 & 0 & t \\
0 & 1 & s & 0 \\
0 & s & -1 & 0 \\
t & 0 & 0 & -2
\end{array}
\right),
\end{equation}
with the spectrum 
\begin{equation}
\left\{-\sqrt{s^2+1},\sqrt{s^2+1},-\sqrt{t^2+4},\sqrt{t^2+4}\right\}.
\end{equation}
\rm Then we can write 
\begin{eqnarray}
\nonumber
\bra{x} \kron \bra{y} X_{t,s} \ket{x} \kron  \ket{y} &=& 
2 |x_1|^2 |y_1|^2 + |x_1|^2 |y_2|^2 - |x_2|^2 |y_1|^2 -2 |x_2|^2 |y_2|^2 \\
&&
+ 2 t Re [ \bar{x}_2 x_1 \bar{y}_2 y_1] + 2 s Re [ \bar{x}_1 x_2 \bar{y}_1 y_2].
\end{eqnarray}
Because in the case of Hermitian matrices the product numerical range 
forms a closed interval, 
we only need to find the upper and the lower bounds for the above expression. We have 
\begin{eqnarray}
\nonumber
\bra{x} \kron \bra{y} X_{t,s} \ket{x} \kron  \ket{y} 
&\leq&
2 |x_1|^2 |y_1|^2 + |x_1|^2 |y_2|^2 - |x_2|^2 |y_1|^2 -2 |x_2|^2 |y_2|^2 \\
&&
+ 2 (t+s)  |x_2| |x_1| |y_2| |y_1|.
\end{eqnarray}
Because $|x_1|^2+|x_2|^2 = 1$, we can put $p=|x_1|^2$ and $q=|y_1|^2$ with $p,q\geqslant 0$. This gives us
\begin{eqnarray}
\nonumber
\bra{x} \kron \bra{y} X_{t,s} \ket{x} \kron  \ket{y} 
&\leq&
2 p q  + p(1-q) - (1-p)q -2 (1-p)(1-q) \\
&&
+ 2 (t+s)  \sqrt{p(1-p)q(1-q)}.
\end{eqnarray}
We want to maximize the above expression under the following constraints: 
$0 \leq p \leq 1 $ and $0 \leq q \leq 1 $.


\begin{figure}[ht!]
\begin{center}
\includegraphics[width=2.4in]{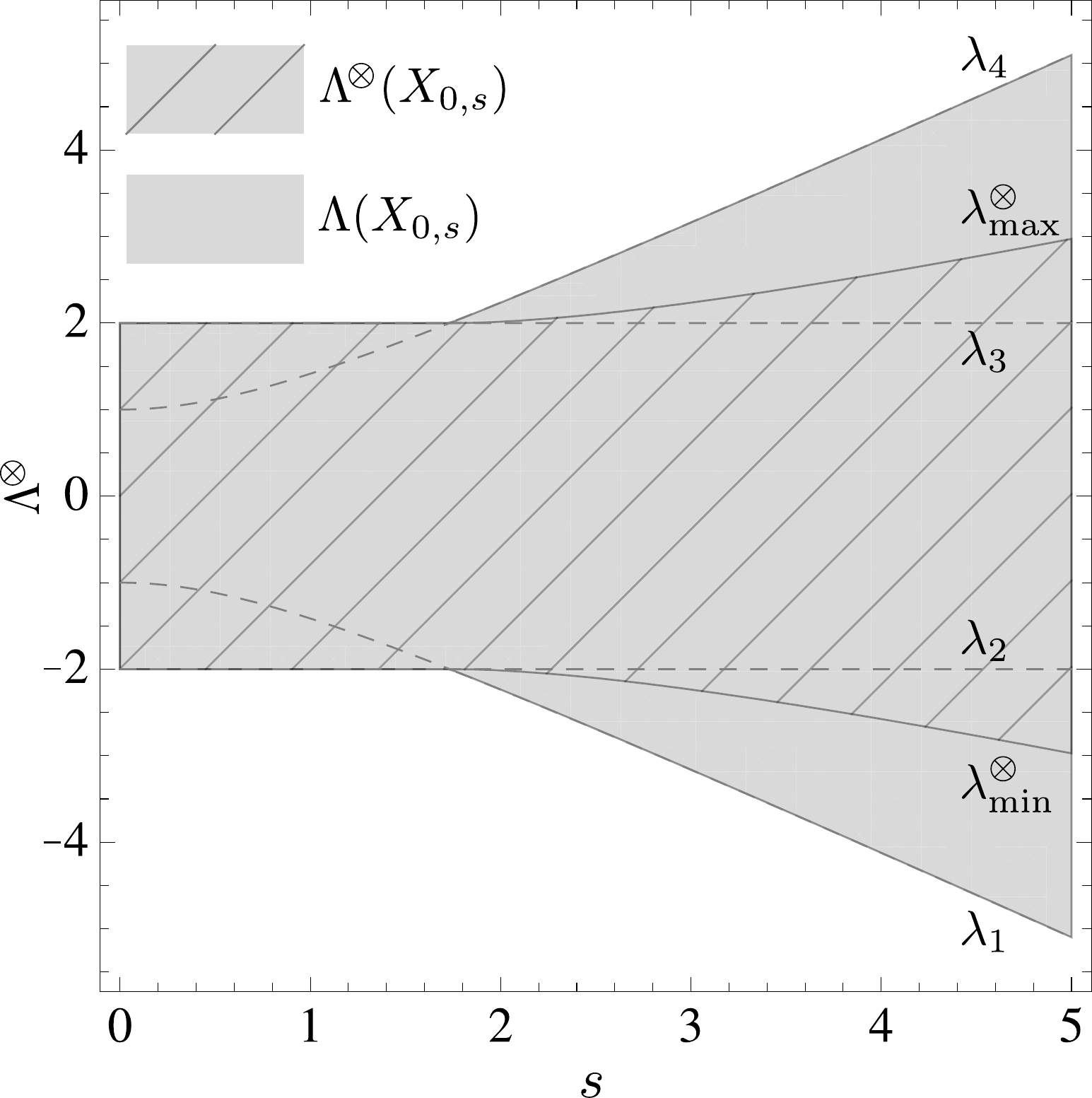}
\caption{Numerical range and product numerical range for matrices $X_{0,s}$,
which belong to the family (\ref{xts}).} 
\label{fig:locmax_vs_max_inter}
\end{center}
\end{figure}

First we analyze the edge. On the edge (one of the variables $p,q$ is 0 or 1)
the square root vanishes, the remaining part is convex and thus the extreme points are
$(p,q) \in \{ (0,0),(0,1),(1,0),(1,1) \}$. Thus the maximum value on the edge is 2.
If we assume that $p,q \notin \{0,1\}$, we have to find zeros of appropriate derivatives.
The extremum  value is 
$\sqrt{t^4+10 t^2+9} / 2 t$  for $t \geq \sqrt{3}$. 
%
The lower estimate is obtained similarly.
Thus the exact formula for the product numerical range reads:
\begin{equation}
\ProductNumRange{X_{t,s}} = [-f(t+s),f(t+s)],
\end{equation}
where
\begin{equation}
f(t) = 
\left \{
\begin{array}{c c c}
 2 & \text{ for } & t \in [0,\sqrt{3}) \\
 \sqrt{t^4+10 t^2+9} / 2 t  
& \text{ for } & t \in [\sqrt{3},\infty) 
\end{array}
\right . .
\end{equation}

Note that the product numerical range depends only on the sum of the parameters
$s$ and $t$, whereas the numerical range depends on the values of both of them.
The minimum and the maximum values in the numerical range and the product
numerical range of the matrix $X_{t,s}$ are compared in
Fig.~\ref{fig:locmax_vs_max_inter}.

Let us consider a more general family of matrices for $t,s \geq 0$
\begin{equation}
Y_{t,s} = 
\left(
\begin{array}{cccc}
a & 0 & 0 & t \\
0 & b & s & 0 \\
0 & s & c & 0 \\
t & 0 & 0 & d
\end{array}
\right).
\end{equation}
For given $a,b,c,d$, one can obtain a similar result as above, but in general the 
formulas are very complex due to the higher number of parameters. 
However, it is easy to obtain the following bound 
\begin{equation}\label{eqn:bound-local-range-example}
 [f(s+t),g(s+t)] \subset \ProductNumRange{Y_{s,t}} ,
\end{equation}
where
\begin{equation}
f(t) = \min \{\min(a,b,c,d), \frac{1}{4} \tr Y_{0,0} - \frac{1}{2} t \},
\end{equation}
and
\begin{equation}
g(t) = \max \{\max(a,b,c,d), \frac{1}{4} \tr Y_{0,0} + \frac{1}{2} t \}.
\end{equation}


\medspace


\subsubsection{A tridiagonal Hermitian matrix}

Consider another family of Hermitian matrices of size four, 
written in the standard product basis,
\begin{equation}
\label{abcfamilydef}
D=
\left[\begin{array}{cccc}
\frac{1}{2}&a&0&0\\
\bar a&\frac{1}{2}&b&0\\
0&\bar b&\frac{1}{2}&c\\
0&0&\bar c&\frac{1}{2}
\end{array}\right],
\end{equation}
where $a$ and $b$ are arbitrary complex numbers and $c=xa$ for some arbitrary real number $x$.

\rm This family was introduced in \cite{SZ09} as a useful example for studying block positivity.
Here we deal with the product numerical range of $D$, but the two concepts are 
closely related, since a Hermitian matrix acting on a bipartite Hilbert space is block positive 
iff its product numerical range belongs to $\R_+$. Following the lines of \cite{SZ09}, 
with some additional effort, one obtains an explicit result 
\begin{equation}
\label{localrange}
\ProductNumRange{D} = \left[\frac{1}{2} -G ,\frac{1}{2}+G \right],
\end{equation}
where
\begin{equation}
\label{Mcaseb}
 G=\frac{1}{4}\left(|a+c|+\sqrt{|a-c|^2+|b|^2}\right)   .
\end{equation}



\subsubsection{Family of isospectral Hermitian operators}

It is instructive to study product numerical range
for a family of Hermitian operators with a fixed spectrum
and varying eigenvectors.
Any unitary $4\times 4$ matrix $U$ may be represented in a canonical form: 
\begin{equation}
U= (V_A\otimes V_B) \;  U_d \;  (W_A\otimes W_B) \; ,
\end{equation}
where $V_A, V_B, W_A, W_B \in U(2)$, while $U_d$ is a unitary matrix of 
size four expressed in the form \cite{KC2001}
\begin{equation}\label{equ:nonlocal-unitary}
	U_d(\alpha_1,\alpha_2,\alpha_3)= \exp(i \sum_{k=1}^3 \alpha_k\sigma_k\otimes\sigma_k) .
\end{equation}
Here $\sigma_k$ denotes the Pauli matrices, and the three real parameters
$\alpha_i$ belong to the interval $\left[0,\frac{\pi}{4}\right]$.

\begin{figure}[ht!]
\begin{center}
\includegraphics[width=3in]{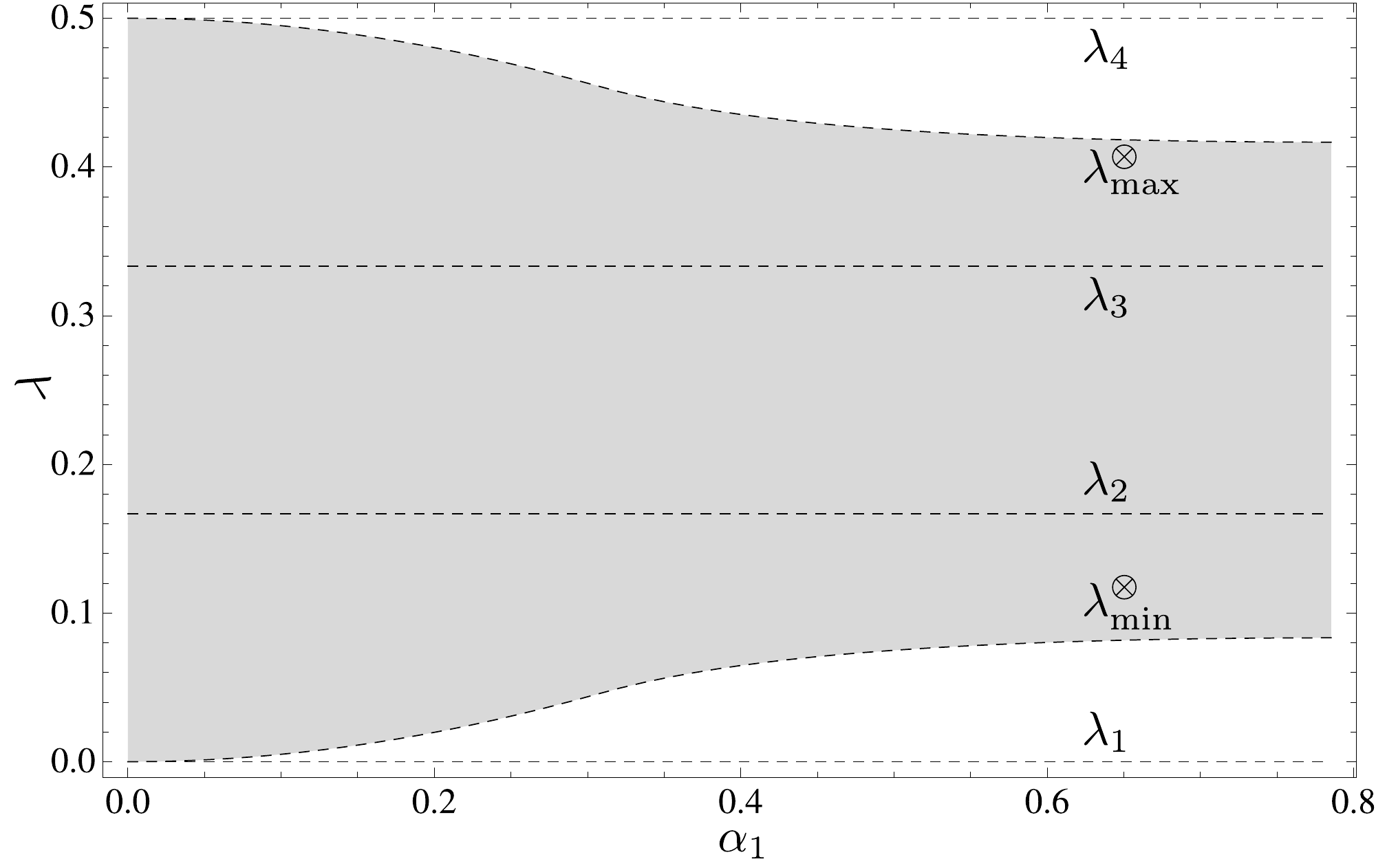}
\caption{Product eigenvalues and product numerical range (gray region)
 of the one-parameter ($\alpha_1$) family of matrices given by Eq.~(\ref{eqn:alpha-ud})
 with eigenvalues $\lambda_1=0, \lambda_2=1/6, \lambda_3=1/3, \lambda_4=1/2$.}
\label{fig:example-canonical}
\end{center}
\end{figure}

Consider a density matrix  obtained from the diagonal matrix 
$E(x_1,x_2,x_3)=\diag(x_1,x_2,x_3,1-x_1-x_2-x_3)$
by a non-local unitary rotation,
\begin{equation}
\label{eqn:alpha-ud}
\rho(\alpha_1,x_i) = U_d E U_d^\dagger,
\end{equation}
with $\alpha_2=\alpha_3=0$.
Figure  \ref{fig:example-canonical} presents the 
dependence of its product numerical range 
as a function of the non-locality phase $\alpha_1$.



\medskip

\subsubsection{Random Hermitian matrices of order four}

As shown in the above examples, the  lower edge 
of the product numerical range of a Hermitian matrix $X$ of order four 
is interlaced between its two smallest eigenvalues,
$\lambda_{\rm min}^{\otimes} \in [\lambda_1, \lambda_2]$.
We have already seen that these bounds can be saturated,
so the exact position of $\lambda_{\rm min}^{\otimes}$
is $X$ dependent. However, following the statistical approach,
one may pose the question how the edge is located with respect to
both eigenvalues for a \emph{random} Hermitian operator.

To analyze this problem we generated numerically a $5 \times 10^5$ random Hermitian matrices
according to the flat (Hilbert--Schmidt) measure
in the set $\Omega$  of normalized density matrices of size $N=4$. 
The joint probability distribution for the eigenvalues reads \cite{SZ04}
\begin{equation}
P(\lambda_1, \lambda_2, \lambda_3 ,  \lambda_4)  \  = \  
\frac{15 !}{3456}
\;  \delta\bigl(1-\sum_{j=1}^4 \lambda_j \bigr)  
 \prod_{i<j}^4 (\lambda_i-\lambda_j)^2 .
   \label{HSmes}
\end{equation}

By construction, the eigenvalues sum to unity,
and this normalization sets the scale.
It is possible to integrate out of the above formula
 any chosen three eigenvalues and obtain an 
explicit probability distribution for the last one. 
For instance the distribution for the smallest eigenvalue has the form 
\begin{equation}
P(\lambda_1)  \ = \ 
             60 (1 - 4 \lambda_1)^{14} \
\Theta(\lambda_1) \;  \Theta(1/4-\lambda_1) \ ,
\end{equation}
where $\Theta(x)$ is the Heaviside function.

\begin{figure}[h!]
\begin{center}
\includegraphics[width=3in]{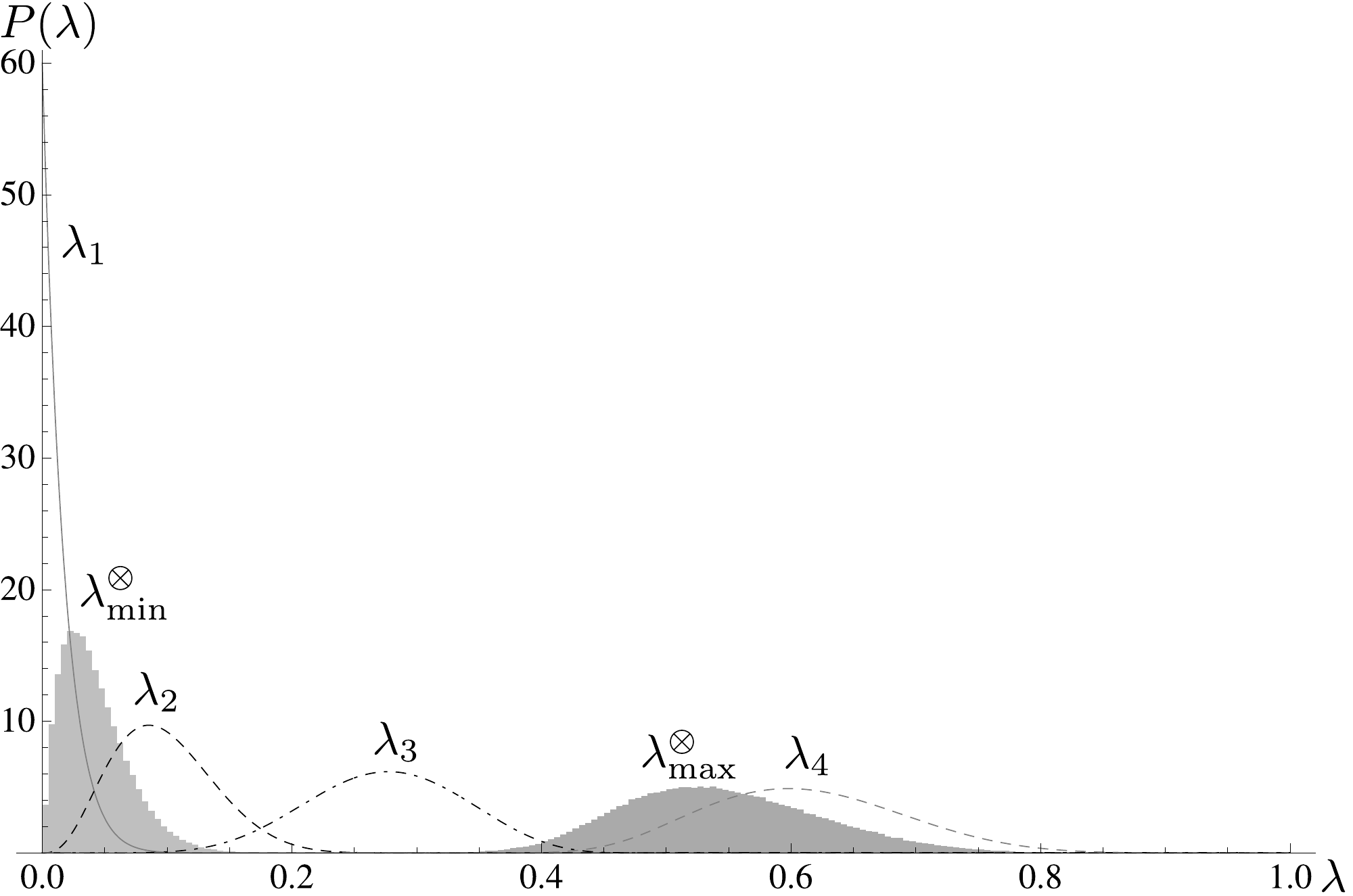}
\caption{Probability density of eigenvalues 
$\lambda_1, \lambda_2, \lambda_3, \lambda_4$ (dashed/dotted lines)
and product values $\lambdaProdMin, \lambdaProdMax$ (dark histograms)
for a random  two-qubit density matrix, generated according to the Hilbert-Schmidt
measure (\ref{HSmes}).
}
\label{fig:lambda-prob}
\end{center}
\end{figure}

\noindent
Figure \ref{fig:lambda-prob} presents the probability 
distributions for ordered eigenvalues, $\lambda_1 \le \lambda_2 
\le \lambda_3 \le \lambda_4$, obtained analytically
by integration of  (\ref{HSmes}).
These distributions are compared with the distributions
$P(\lambdaProdMin)$ and $P(\lambdaProdMax)$ obtained numerically.
As follows from (\ref{herm22}) 
$\lambda_{\rm min}^{\otimes}$
is located between the two smallest eigenvalues,
while $\lambda_{\rm max}^{\otimes}$
is interlaced by the two largest eigenvalues
$\lambda_3$ and $\lambda_4$. 
Note that the histogram is not symmetric
with respect to the change $\lambda_1 \leftrightarrow \lambda_4$
and $\lambda_{\rm min}^{\otimes} \leftrightarrow \lambda_{\rm max}^{\otimes}$,
since the eigenvalues are ordered, so 
the mean distance of the smallest eigenvalue to zero is 
smaller than the mean distance of the largest eigenvalue to unity.



\subsection{Non--Hermitian case and Multipartite operators}
\label{sec:multipartite}
The above analysis can be  extended in a natural way for 
Hilbert spaces with $m$-fold tensor product structure, 
used to describe quantum systems consisting of $m$ subsystems,
\begin{equation}
\mathcal{H}_N \ = \  \mathcal{H}_{n_1} \otimes \cdots \otimes \mathcal{H}_{n_m},
\end{equation}
with $N=n_1\dots n_m$.
In the case of an operator $X$ acting on this space,
its product numerical range consists of all expectation values
$\langle \psi_{\rm prod}|X|\psi_{\rm prod}\rangle$
among pure product states,
$|\psi_{\rm prod}\rangle=|\phi_1\rangle \otimes \dots \otimes 
|\phi_m\rangle$.

\begin{figure}[htb]
\begin{center}
	\renewcommand{\thesubfigure}{\ }
	\subfigure[$n=1$]{\includegraphics[width=0.2\textwidth]{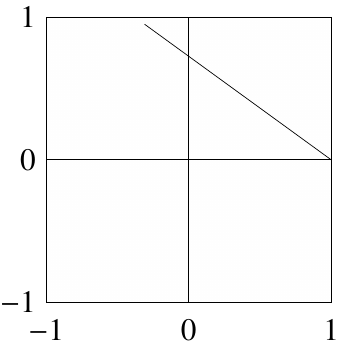}}
	\subfigure[$n=2$]{\includegraphics[width=0.2\textwidth]{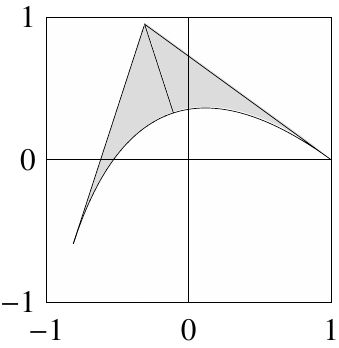}}
	\subfigure[$n=3$]{\includegraphics[width=0.2\textwidth]{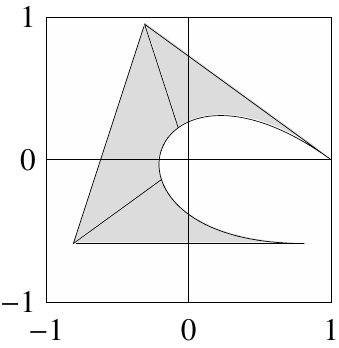}}
	\subfigure[$n=4$]{\includegraphics[width=0.2\textwidth]{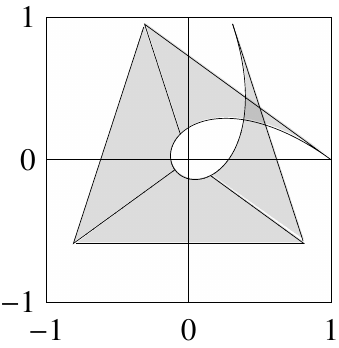}}\\
	\subfigure[$n=5$]{\includegraphics[width=0.2\textwidth]{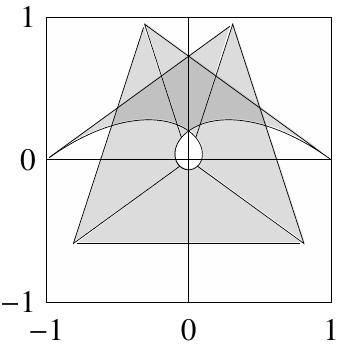}}
	\subfigure[$n=6$]{\includegraphics[width=0.2\textwidth]{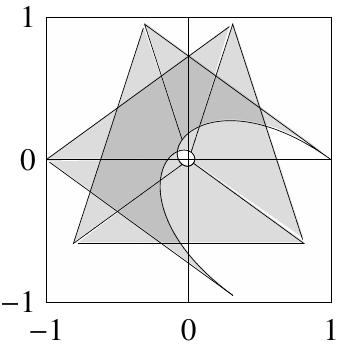}}
	\subfigure[$n=7$]{\includegraphics[width=0.2\textwidth]{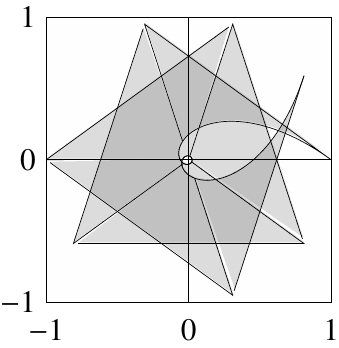}}
	\subfigure[$n=8$]{\includegraphics[width=0.2\textwidth]{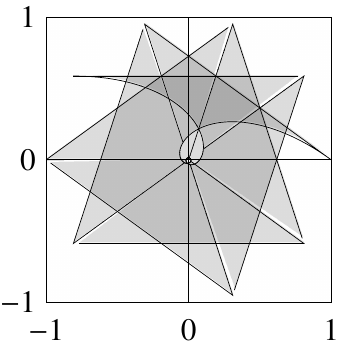}}
\end{center}
\caption{\label{listaobrazkow} Product numerical range of 
$U^{\otimes n}$ for $U$ specified in \eqref{generalA} with $\phi= 3\pi / 5$
and $n=1,\ldots,8$.}
\end{figure}

If the number $m$ of subsystems is larger than two,
there exist operators for which 
product numerical range forms a set which is not simply connected
\cite{thomas08significance,PGMSCZ10}.
 In fact the genus of this set can be greater than one.
 To show an illustrative example, we consider 
a unitary matrix of size two
\begin{equation}
\label{generalA}
 U=\left[
\begin{array}{cc}
 1&0\\
0&e^{i\phi}
\end{array}
\right].
\end{equation}
The product numerical range of $U^{\otimes n}$ can
 be found analytically for any integer $n$
by applying an extension of the formula
(\ref{rangetensorpr}) to multipartite systems.
Numerical range of $U$ forms an interval $I$ joining
the complex eigenvalue $e^{i\phi}$ with the unity.
Thus to find $\ProductNumRange{U^{\otimes n}}$,
it suffices to compute the $n$-fold Minkowski power
of the interval $I$ on the complex plane. More explicitly, $\ProductNumRange{U^{\otimes n}}$ consists of all the points $z_1z_2\ldots z_n$, where $z_i=1-\lambda_i+\lambda_ie^{i\phi}$ and $\lambda_i\in\left[0,1\right]$ for all $i\in\left\{1,2,\ldots,n\right\}$. Let us denote by $f\left(\alpha\right)$ the modulus $\left|z\right|$ of $z=1-\lambda+\lambda e^{i\phi}$ as a function of the phase $\alpha:=\textnormal{Arg}\left(z\right)$. Obviously, $f$ is a convex function of $\alpha$. One can relatively easy get an explicit expression for $f$,
\begin{equation}\label{exprf}
 f\left(\alpha\right)=\frac{\cos\left(\phi/2\right)}{\cos\left(\alpha-\phi/2\right)}.
\end{equation}
Thus the numbers $z$ of the form $z=1-\lambda+\lambda e^{i\phi}$, $\lambda\in\left[0,1\right]$ have a parametrization $\alpha\mapsto e^{i\alpha}f\left(\alpha\right)$ with $\alpha\in\left[0,\phi\right]$ and $f$ given by formula \eqref{exprf}. Because of the convexity of $f$, for a fixed $\textnormal{Arg}\left(z_1z_2\ldots z_n\right)$, the minimum of $\left|z_1z_2\ldots z_n\right|$ is attained when $z_1=z_2=\ldots=z_n$.
The resulting curve marks the border of $\ProductNumRange{U^{\otimes n}}$ and has a parametrization
\begin{equation}
\left[0,\phi\right]\ni\alpha\mapsto 
e^{i n\alpha}
\left(\frac{\cos\left({\phi/2}\right)}
{\cos\left(\alpha-{\phi/2}\right)}\right)^n.
\end{equation}
The remaining parts of the border of $\ProductNumRange{U^{\otimes n}}$ are included in the $n$ segments $\left\{\left[e^{i\left(k-1\right)\phi},
e^{ik\phi}\right]\right\}_{k=1}^n$. This follows because the maximum of $\left|z_1z_2\ldots z_n\right|$ for a fixed $\beta=\textnormal{Arg}\left(z_1z_2\ldots z_n\right)$ is attained for $z_1=z_2=\ldots=z_{k-1}=e^{i\phi}$, $z_k=e^{i\left(\beta-k\phi\right)}$ and $z_{k+1}=z_{k+2}=\ldots=z_n=1$, where $k=\lfloor\beta/\phi\rfloor$. 
In Figure~\ref{listaobrazkow} we choose $\phi=\frac{3\pi}{5}$ and plot the product 
numerical ranges of $U^{\otimes n}$ for $n=1,2,\ldots,8$.

\begin{figure}
 [htb]
\begin{center}
 \includegraphics[scale=0.55]{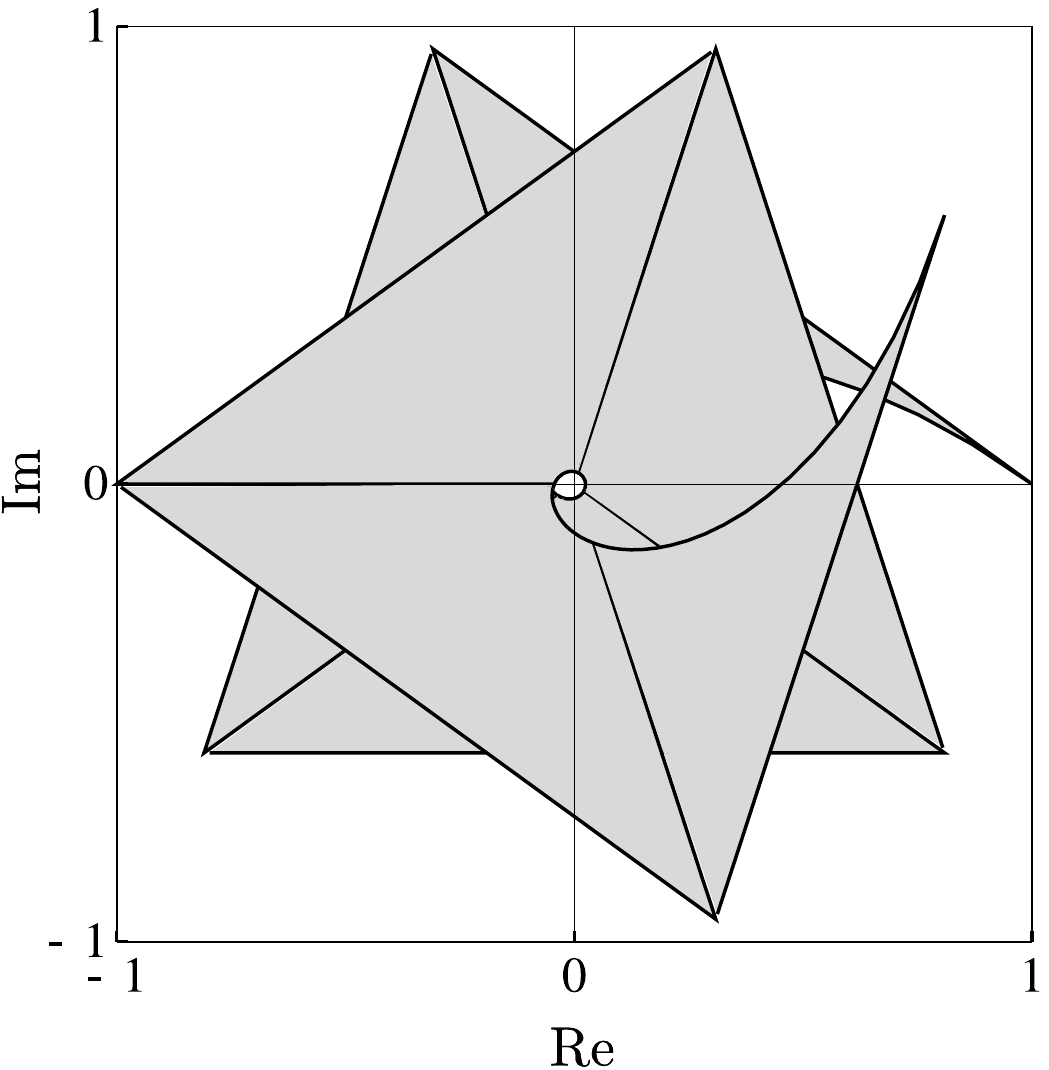}
\end{center}
\caption{\label{figlocrange2product} Product numerical range of 
 $U^{\otimes 7}$ (plotted in gray) forms a set of genus $2$.
 The matrix $U$ is given by Eq. \eqref{generalA} 
 with $\phi= 3\pi / 5$.}
\end{figure}

\rm Observe that if $0$ does not belong to the numerical range of $U$,
it does not belong to the product numerical range of $U^{\otimes n}$.
 Hence if for sufficiently
large exponent $n$ the product range 'wraps around' zero, the set 
$\ProductNumRange{U^{\otimes n}}$ is not simply connected.

As one may notice in Figure~\ref{listaobrazkow}, it is possible to construct a
tensor product of~operators such that its product numerical range has genus two.
If we magnify picture number seven from Figure~\ref{listaobrazkow}, it becomes
evident that the genus of~$\ProductNumRange{U^{\otimes 7}}$ is~equal to two (cf.
Figure~\ref{figlocrange2product}). Observe that if $n$ is further increased, the
genus of~$\ProductNumRange{U^{\otimes n}}$ is not smaller than one, although the
size of~the~hole around $z=0$ shrinks exponentially fast. More precisely, the distance 
between the set $\Lambda\left(U\right)$ and
zero is $\cos\left(\phi/2\right)$, which implies that the distance
between $\Lambda^{\otimes}\left(U^{\otimes n}\right)$ and
zero equals $[\cos\left(\phi/2\right)]^n$ for arbitrary $n$.

In general, finding the product numerical range of a non-Hermitian 
operator  without the tensor product structure is 
not a simple task. However, in the special case of  a normal
operator $X$, which can be diagonalized by product of unitary matrices,
a useful parameterization of its product numerical range
was described in \cite{PGMSCZ10}.

\section{Separable numerical range}
\label{sec:separable}

Consider a tensor product Hilbert space
$\mathcal{H}_N =\mathcal{H}_K \otimes \mathcal{H}_M$
and the set $\Omega$ of all normalized states acting on it, 
$\rho \in \Omega \Leftrightarrow \rho=\rho^{\dagger}, \rho\ge 0,
 {\rm Tr}\rho=1$.
One distinguishes its subset $\Omega_{\rm sep}$ 
of {\it separable states}, i.e. states that
can be represented as a convex combination of product states,
 
 \begin{equation}
\rho \in \Omega_{\rm sep} 
\ \Leftrightarrow \
\rho \in \Omega {\rm \quad and \quad}
\rho=\sum_i p_i \; \rho_i^{(K)} \otimes \rho_i^{(M)}
   \label{separ}
\end{equation}
Here positive coefficients $p_i$ form a probability vector, 
while  $\rho_i^{(K)}$  and $\rho_i^{(M)}$
denote arbitrary states acting on 
$\mathcal{H}_K$ and $\mathcal{H}_M$, respectively. 
Any state $\rho$ which cannot be represented in the above form
is called {\it entangled} \cite{BZ06}.
 Hence this definition depends on the 
particular choice of the tensor product structure, 
$\mathcal{H}_N =\mathcal{H}_K \otimes \mathcal{H}_M$.

Observe that Definition \ref{lnr}
of the product numerical range of an operator $X$
acting on  $\mathcal{H}_K \otimes \mathcal{H}_M$
can be formulated as 
\begin{equation}
  \ProductNumRange{X} = \left\{ {\rm Tr} X\rho
: \rho= |\psi_A\rangle \otimes |\psi_B\rangle
        \langle \psi_A| \otimes \langle \psi_B|\right\} .
   \label{lnrbis}
\end{equation}
It is then natural to introduce an analogous definition 
of {\it separable numerical range}
 \begin{equation}
  \Lambda^{\rm sep} (X) := \left\{ {\rm Tr} X\rho
: \rho \in \Omega_{\rm sep} \right\}.
   \label{sepnr}
\end{equation}
Since any product state is separable, the product numerical range
forms a subset of the separable numerical range,
$\ProductNumRange{X} \subset \Lambda^{\rm sep} (X)$.
By definition, the set $\Omega_{\rm sep}$ of separable states is convex.
This fact allows us to establish a simple relation between both sets.

\begin{proposition}
 \label{prop:prodsep}
Separable numerical range forms the convex hull
of the product numerical range, 
 $$ \Lambda^{\rm sep} (X) = {\rm co} (\ProductNumRange{X})  $$
\end{proposition}

\proof{Assume that $\lambda \in 
{\rm co} (\ProductNumRange{X})$, so it can be represented as 
a convex combination of points
belonging to the product numerical range,  $\lambda=\sum_i p_i \lambda_i$.
 Taking the convex combination of the corresponding product states
 $|\phi_i\rangle=|\psi_i^A\rangle  \otimes |\psi_i^B \rangle$ 
we get a separable mixed state
$\rho=\sum_i p_i  |\phi_i\rangle \langle \phi_i|$
such that ${\rm Tr}X \rho=\lambda$.
 A similar reasoning shows that if
 $\lambda \notin {\rm co} (\ProductNumRange{X})$
there is no separable state $\rho$ such that
${\rm Tr}X \rho=\lambda$.
 }
\halmos

Following \cite{PGMSCZ10} one can note that if $A$ or $B$ is normal then
$\Lambda^{\rm sep}(A\otimes B)=\Lambda(A\otimes B)$.

\begin{figure}
\centering
\begin{tabular}{cc}
\begin{overpic}[width=0.30\textwidth]{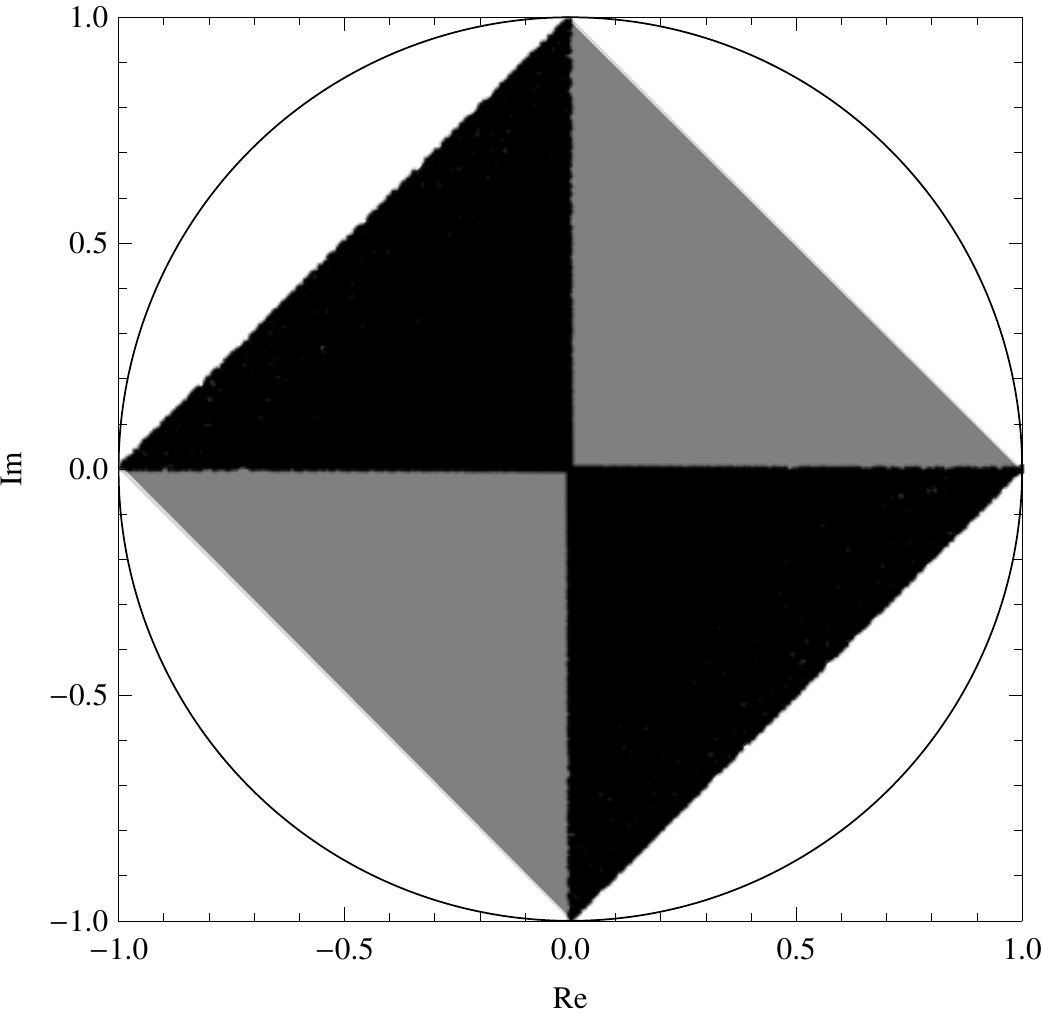}
\put(80,12){\scriptsize $\alpha=0$}
\end{overpic}
\begin{overpic}[width=0.30\textwidth]{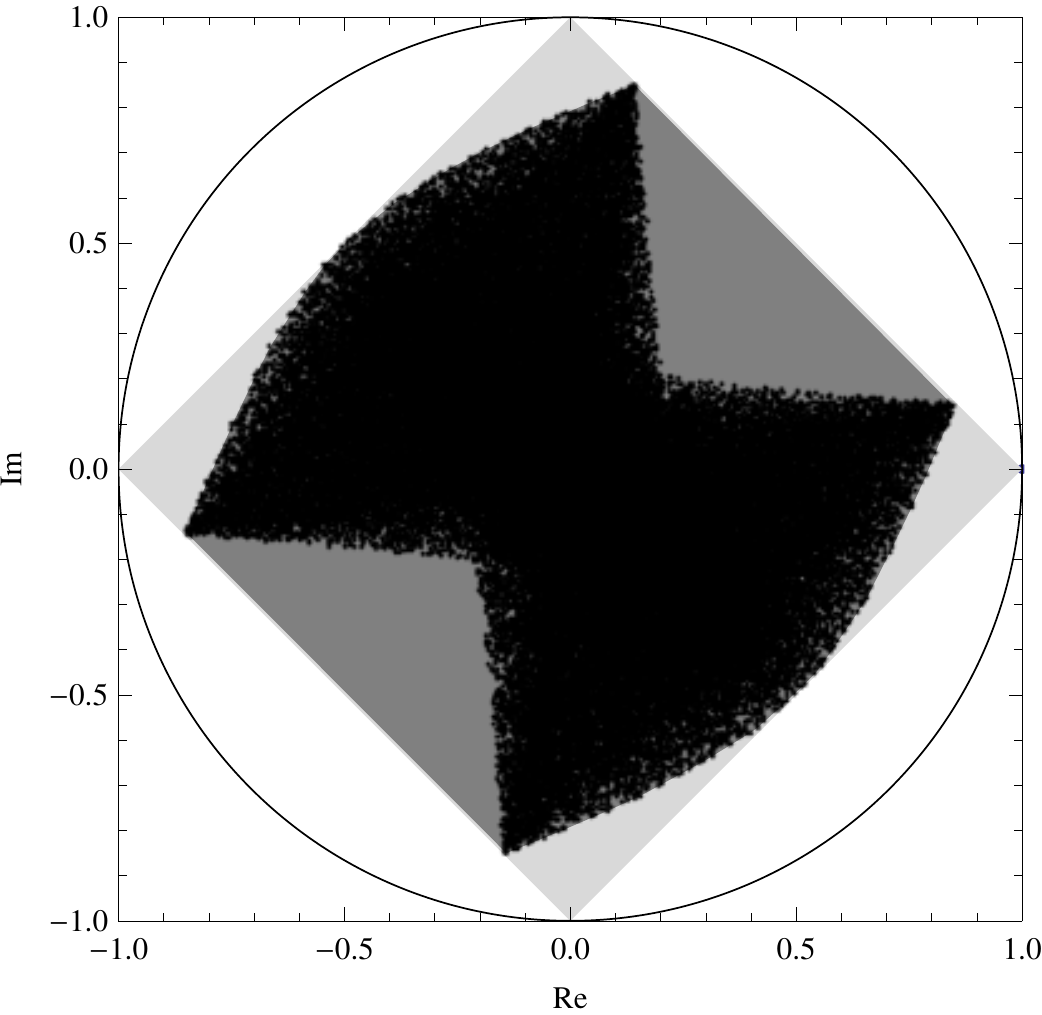}
\put(80,12){\scriptsize $\alpha=\frac{\pi}{8}$}
\end{overpic}\\
\begin{overpic}[width=0.30\textwidth]{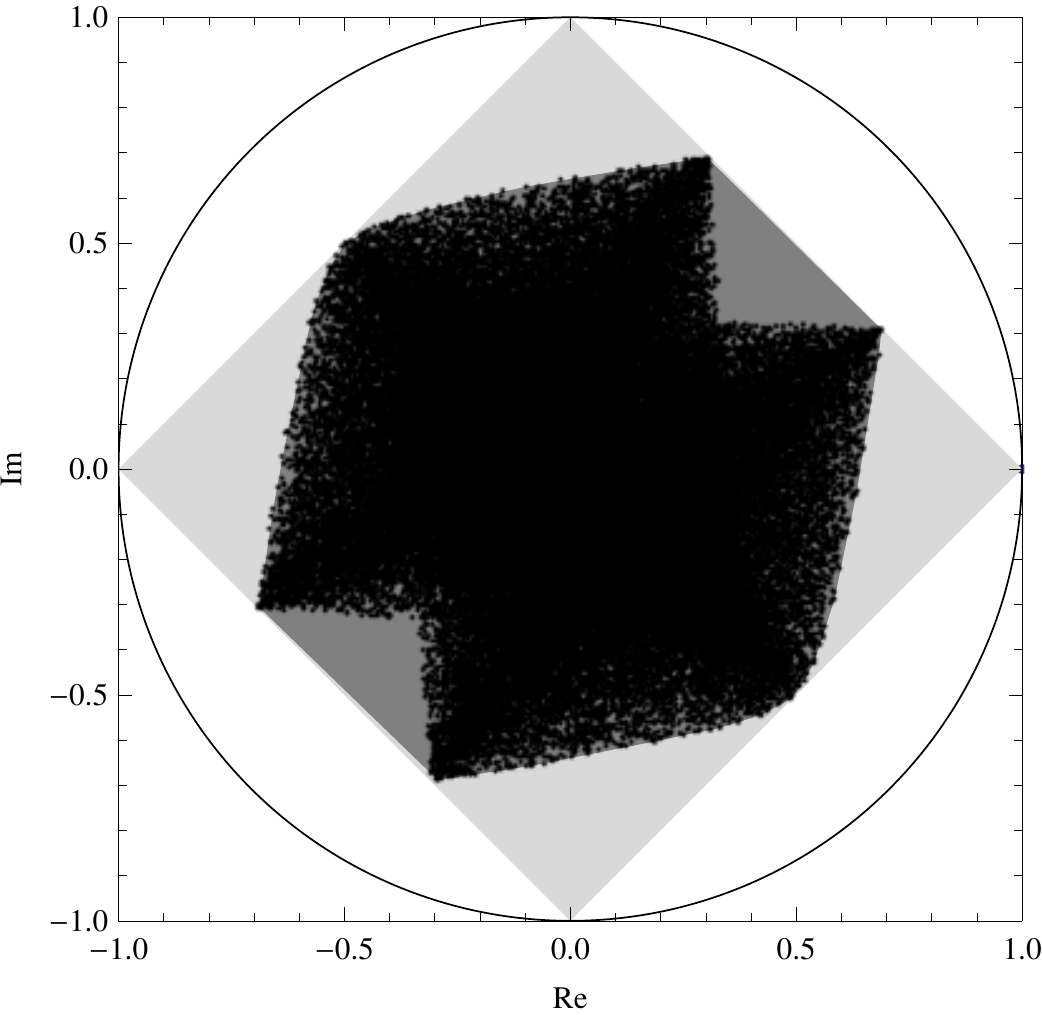}
\put(80,12){\scriptsize $\alpha=\frac{3 \pi}{16}$}
\end{overpic}
\begin{overpic}[width=0.30\textwidth]{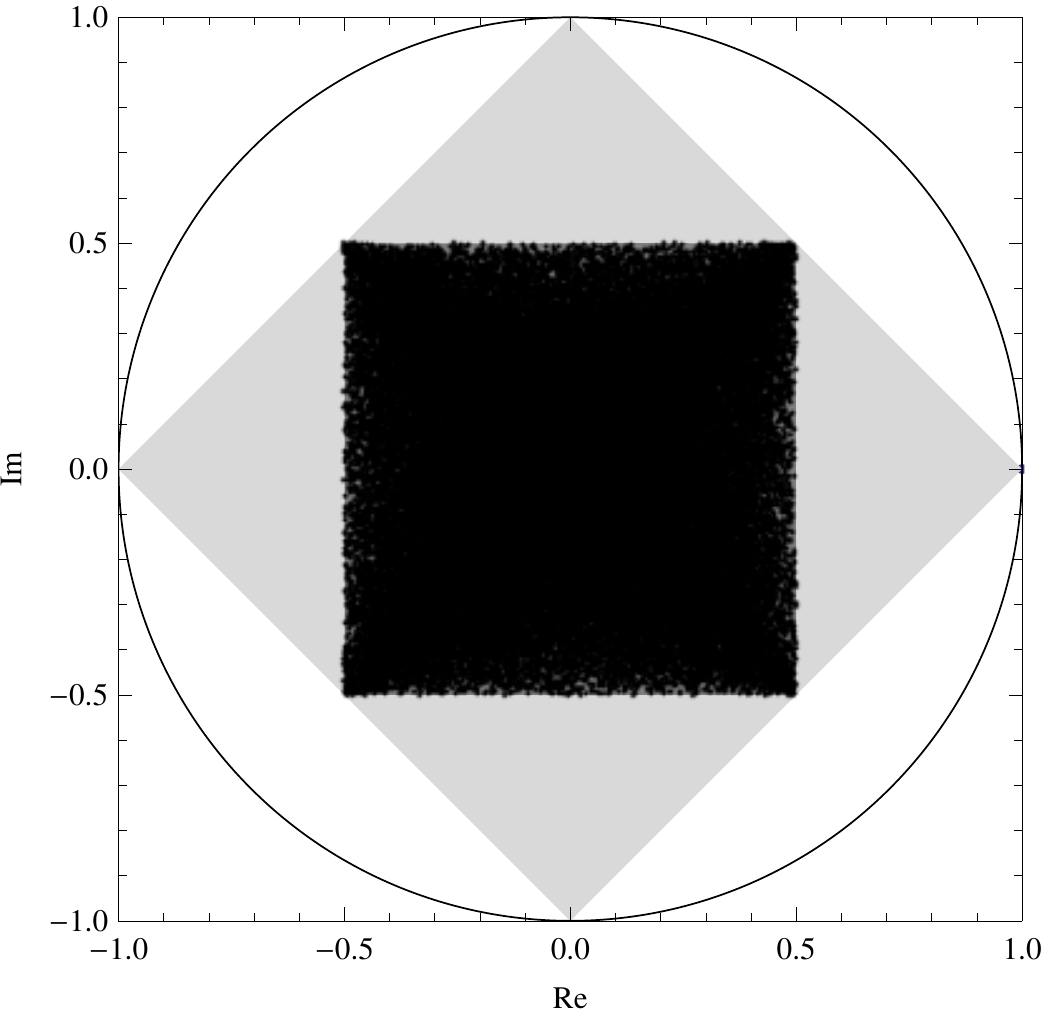}
\put(80,12){\scriptsize $\alpha=\frac{\pi}{4}$}
\end{overpic}
\end{tabular}
\caption{Numerical range (light gray), separable numerical 
range (dark gray) and product numerical range (black dots obtained by 
random sampling) of family of matrices 
$X_\alpha=U_d(\alpha,0,0)\cdot\diag(i, -1, -i, 1)\cdot U_d(\alpha,0,0)^\dagger$,
where $U_{\alpha}$ is given by Eq.~(\ref{equ:nonlocal-unitary}) for $\alpha=0,\pi/8, 3\pi/16, \pi/4$.
In the case of $\alpha=0$ the eigenvectors of $X$ form orthonormal 
canonical basis and $X$ is normal therefore $\Lambda^{\rm sep}(X)=\Lambda(X)$. 
In the case of $\alpha=\pi/4$ all eigenvectors of $X$ are maximally entangled states 
and $\Lambda^{\rm sep}(X)=\ProductNumRange{X}$.
}
\label{fig:separable}
\end{figure}

Since product numerical range of a Hermitian operator
forms an interval, in this case 
the separable and product numerical ranges do coincide.
This is not the case in general.
A typical example is shown in Figs.~\ref{fig:separable}b and \ref{fig:separable}c,
in which the separable numerical range forms a proper
subset of the standard numerical range and includes the
product numerical range as its proper subset.

Consider, for instance a unitary matrix  $U$ of size $4$
with a non-degenerate spectrum. Its numerical
range is then formed by a quadrangle inscribed
into the unit circle. If all eigenvectors of this matrix 
are entangled, the product numerical range of $U$
does not contain any of its eigenvalues.
In a generic case  $\ProductNumRange{U}$
is not convex and it forms a proper subset of
$\Lambda^{\rm sep} (U)$ -- see Fig.~\ref{fig:separable}.

\subsection{$k$--Entangled numerical range}\label{kentsec}
Any pure state in a $N=KM$ dimensional bipartite Hilbert space
 can be represented by its \emph{Schmidt decomposition},
\begin{equation}
   |\psi\rangle   =   \sum_{i=1}^K \sum_{j=1}^M 
   A_{ij} |i\rangle \otimes |j\rangle 
      =  
  \sum_{i=1}^{K} \sqrt{\mu_i}\; |i' \rangle \otimes |i''\rangle.
  \label{Aij}
  \end{equation}
We the have assumed here that $K\le M$
and denoted a suitably rotated product basis by $|i' \rangle \otimes |i''\rangle $.
The eigenvalues $\mu_i$ of a positive matrix $AA^{\dagger}$ are called the 
\emph{Schmidt coefficients} of the bipartite state  $|\psi\rangle$. The 
normalization condition $|\psi|^2=\langle \psi|\psi \rangle=1$ implies that 
$||A||^2_{\rm HS}=\tr AA^{\dagger}=1$, so the Schmidt coefficients 
$\mu_i$ form a probability vector -- see e.g. \cite{BZ06}.
 
The state $|\psi\rangle$  is separable iff the $K \times M$ matrix of 
coefficients $A$ is of rank one, so the corresponding vector of the Schmidt 
coefficients is pure. 
A given mixed state $\rho$ is called separable  
if it can be represented as a convex combination of
product pure states.
This notion can be generalized in a natural way, and 
in the theory of quantum information \cite{TH00}
once considers set $\Omega^{({\rm k})}$
of  states  which can be decomposed
into a convex combination of states
with the Schmidt number not larger than $k$. 
%
%
In symbols,   
$\rho=\sum_{j=1} p_j
\ketbra{\phi_j}{\phi_j},$ with all vectors 
$\ket{\phi_j}=\sum_{i=1}^k \xi_i \ket{\psi^A_i}\otimes \ket{\psi^B_i}$
of Schmidt rank at most $k$.
We may choose $k$ to be $1, \dots, K$,
where $K=M$ denotes the dimensionality of each subsystem.
By definition, $\Omega^{(1)}= \Omega_{\rm sep}$
represents the set of separable states, while
$\Omega^{({\rm K})}= \Omega$ denotes the entire
set of mixed quantum states.

Making use of the definition of the subset 
 $\Omega^{({\rm k})}$ of the set of all states
in (\ref{rest1}) one obtains 
an entire hierarchy of restricted
numerical ranges denoted by $\Lambda^{(k)}$.
As the elements of  $\Omega^{({\rm k})}$
are called $k$-entangled states \cite{SSZ09}, 
the  set $\Lambda^{({\rm k})}(X)$
will be referred to as numerical range restricted to $k$--entangled states.

For $k=1$ one has $\Omega^{(1)}=\Omega_{\rm sep}$
so in this case one obtains the separable numerical range,
$\Lambda^{(1)}=\Lambda^{\rm sep}$.
Note that in this convention a $1$-entangled state means a separable state.
In the other limiting case $k=K$, $\Omega_K=\Omega$
and one arrives at the standard numerical range,
$\Lambda^{(K)}=\Lambda$.
The following chain of inclusions
$\Lambda^{\otimes} \subset \Lambda^{(1)} \subset \Lambda^{(2)}
\subset \cdots \subset \Lambda^{({\rm K})}=\Lambda$
holds by construction. This implies inequalities between the corresponding restricted numerical radii,
   $r^{\otimes} \le r^{(1)} \le r^{(2)}  \le \dots, \le r^{(K)} = r$.


\section{Applications in quantum information theory}
\label{sec:qinfo}

In this section we link various problems in the theory of quantum information
processing which have one thing in common:
they can be analyzed using the restricted 
numerical range  or related notions.

\subsection{Block positive matrices and entanglement witnesses}
Let us start by recalling the standard definition of block-positivity \cite{BZ06}.
A Hermitian matrix $X$ acting on the tensor product Hilbert space, ${\cal H}_N={\cal H}_M 
\otimes {\cal H}_K$, is called \emph{block positive}, if 
it is positive on all product states.
Making use of the notation  introduced in Sec. \ref{sec:product1}
this property reads, $\lambdaProdMin(X) \ge 0$. 
Therefore checking if a given Hermitian matrix is block-positive 
is equivalent 
to showing that its product numerical range forms a subset of $[0,\infty)$.

Block positive matrices arise in a characterization of positive quantum maps by the theorem 
of Jamio{\l}kowski \cite{Ja72}. A map $\Phi$ taking operators on $\mathcal{H}_K$ to 
operators on $\mathcal{H}_M$, is called \emph{positive},
 if it maps positive operators to positive operators.
Let $|\Psi_+\rangle\langle\Psi_+|$ denote the orthogonal projection onto the maximally
 entangled state $|\Psi_+\rangle=\frac{1}{\sqrt{K}}\sum_{i=1}^K|i\rangle|i\rangle$
acting on $\mathcal{H}_K\otimes\mathcal{H}_K$.
The Jamio{\l}kowski theorem states that $\Phi$ is positive iff
the corresponding dynamical matrix (Choi matrix \cite{Cho75a}),
$D_{\Phi}=\left(\Phi\otimes\mathbbm{1}\right)|\Psi_+\rangle\langle\Psi_+|$, is block positive.
This leads us to the following characterization of positive maps in terms of the product
numerical range of $D_{\Phi}$,
\begin{proposition}
\label{charposmaps}
 Let $\Phi$ be a linear map taking operators on $\mathcal{H}_K$ to operators on $\mathcal{H}_M$. Then
\begin{equation}\label{eqcharposmaps}
 \Phi\textrm{ is positive } 
	\Leftrightarrow \ \ProductNumRange{D_{\Phi}}\subset [0,\infty).
\end{equation}
\end{proposition}
That is, the product numerical range of $D_{\Phi}$ has to be contained in the
positive semiaxis in order for $\Phi$ to be positive.
As discussed in Section \ref{sec:separable},
for any Hermitian $D$, its product and separable numerical ranges do coincide.
Consequently, positivity of $\Phi$ can be formulated with $\Lambda^{\rm sep}(D_{\Phi})$. The positivity condition reads: ${\rm Tr} D_{\Phi}\rho\ge 0$ for any separable $\rho$. This is the same as $\Lambda^{\rm sep}\left(D_{\Phi}\right)\subset\left[0;+\infty\right)$.

We recall that a map $\Phi$ is called $k$-{\emph positive}
 if $\Phi\otimes\mathbbm{1}_k$ is a positive map.
If this is the case for arbitrary $k\in\mathbbm{N}$
the map is called \emph{completely positive}.
 The famous theorem  by Choi \cite{Cho75a} concerning 
completely positive maps can be expressed in a similar manner, 
\begin{proposition}
\label{charcomplposmaps}
 Let $\Phi$ be a linear map taking operators on $\mathcal{H}_K$
 to operators on $\mathcal{H}_M$. Then
\begin{equation}\label{eqcharcomplposmaps}
 \Phi\textrm{ is completely positive } 
	\Leftrightarrow \ \LambdaRm\left(D_{\Phi}\right)\subset [0,\infty).
\end{equation}
\end{proposition}
The difference is that \eqref{eqcharposmaps} refers to the product numerical range 
of $D_{\Phi}$ whereas \eqref{eqcharcomplposmaps} concerns the standard numerical range.
Note that $\LambdaRm\left(D_{\Phi}\right)\subset [0,\infty)$ is just another way of 
writing that $D_{\Phi}$ is a positive operator.

Positive maps find a direct application in the theory of quantum information
due to a theorem by the Horodecki family \cite{HHH96a}:
a state $\sigma$ of a bipartite system is separable iff
$(\Phi \otimes {\mathbbm 1}) \sigma \ge 0$ for any positive map $\Phi$.
In the opposite case, the state $\sigma$ is entangled. 

The above results explain recent interest in characterization
of the set of positive maps. A block positive matrix $W:=D_{\Phi}$
which corresponds to a map which is positive but not completely positive,
is called an \emph{entanglement witness}, 
since it can be used to detect quantum entanglement.
As discussed in sec. \ref{sec:separable},
product and separable numerical ranges  coincide for  Hermitian operators.
Thus the set of entanglement witnesses consists of Hermitian operators $W$ 
such that ${\rm Tr}W\rho \ge 0$ for all separable $\rho$
and there exists an entangled state $\sigma$ such that
 ${\rm Tr}W\sigma < 0$. 
The set of separable quantum states can thus be characterized by a suitably chosen 
set of entanglement witnesses. Such an approach was advocated 
in a recent work by Sperling and Vogel \cite{SV09}, 
in which various methods for obtaining the minimal product value $\lambdaProdMin$
of Hermitian matrices were analyzed.

Bound \ref{hermbounds} implies that the spectrum of an entanglement witness
for any state of a $K \times K$ system 
has at most $(K-1)^2$ negative eigenvalues, in accordance with recent results of Sarbicki
\cite{Sa08}. In the simplest case of $K=2$, one recovers the known statement that 
any non-trivial entanglement witness in the two-qubit system has exactly one negative eigenvalue 
\cite{STV98}.

Our study of product numerical range of a Hermitian operator can thus be
directly applied to the positivity problem. For instance, consider the family of
one-qubit maps described by the dynamical matrix $D=D(a,b,c)$ defined in
(\ref{abcfamilydef}). It is clear that these matrices are block positive iff $G
\le 1/2$. Therefore the expression (\ref{Mcaseb}) for $G=G(a,b,c)$ gives us
explicit constraints under which the map corresponding to $D(a,b,c)$ is
positive. If this map is not completely positive, the matrix $D$ can be used as a
witness of quantum entanglement.

In the above case corresponding to maps acting on  $2$ dimensional  
Hilbert space ${\cal H}_2$ any $2$--positive map is completely positive. 
This is a consequence of the theorem of Choi \cite{Cho75a}, 
which implies that if a map acting on $K$ dimensional Hilbert space
is $K$ positive, it is also completely positive. 
Thus  for maps acting on a $K$--dimensional system,
it is interesting to study  $k$--positivity
for $k=1$ (equivalent to positivity),
$k=2,\dots K-1$ and $k=K$ (complete positivity).
In general $k$-block positive matrices
are related to $k$--positive maps. 
We are thus  in a position to formulate the generalized
Jamio{\l}kowski--Choi theorem \cite{Ra07, SSZ09}
making use of the concept of the restricted numerical range.

\begin{proposition}
\label{charcomplposkmaps}
 Let $\Phi$ be a linear map taking operators on $\mathcal{H}_K$
 to operators on $\mathcal{H}_M$. Then
\begin{equation}
\label{eqcharposkk}
 \Phi\textrm{ is $k$--positive } 
	\Leftrightarrow \ \Lambda^{(k)}\left(D_{\Phi}\right)\subset [0,\infty).
\end{equation}
\end{proposition}

\noindent As we explain in the next section, a special case of Proposition \ref{charcomplposkmaps} for $k=2$ is of relevance to the distillability problem for quantum states. 

\subsection{$n$--copy distilability of a quantum state}

It has been known for a long time \cite{diVicenzo00} that bi-partite states with distillable entanglement are closely related to $2$-positive maps and hence to $2$-block positive operators (cf. also \cite{Cl05,SSZ09}). The precise relation between distillability and $2$-block positivity is the following. Let $\rho$ be an arbitrary state on a bipartite space $\mathcal{H}_N=\mathcal{H}_K\otimes\mathcal{H}_M$. Assume that we allow only LOCC operations on a single copy of $\rho$. The state can be distilled into a maximally entangled state only if the partial transpose $\left(\1\otimes T\right)\rho$ is \emph{not} a $2$-block positive operator, i.e. it is not positive on states with Schmidt rank $2$. Otherwise, $\rho$ is one-copy undistillable. Writing this in terms of $k$-entangled numerical ranges (cf. Section \ref{kentsec}), we get the following proposition.
\begin{proposition}\label{1copydistill}
 A state with a density matrix $\rho$ on a bi-partite space $\mathcal{H}_N=\mathcal{H}_K\otimes\mathcal{H}_M$ is one-copy undistillable if and only if the $2$-entangled numerical range of its partial transpose is contained in the nonnegative semiaxis, $\Lambda^{\left(2\right)}\left(\left(\1\otimes T\right)\rho\right)\in\left[0;+\infty\right)$.
\end{proposition}
If a state $\rho$ turns out to be one-copy undistillable, it is still possible that a number of copies of $\rho$ can be used for entanglement distillation. Proposition \ref{1copydistill} is easily generalized to that situation.
\begin{proposition}\label{ncopydistill}
 Let $\rho$ correspond to a state on a bi-partite space $\mathcal{H}_N=\mathcal{H}_K\otimes\mathcal{H}_M$. For any integer $n$, the state is $n$-copy undistillable if and only if the $2$-entangled numerical range of $\left(\1\otimes T\right)\rho^{\otimes n}$ is contained in the nonnegative semiaxis, $\Lambda^{\left(2\right)}\left(\left(\1\otimes T\right)\rho^{\otimes n}\right)\in\left[0;+\infty\right)$.
\end{proposition}
The symbol $\Lambda^{\left(2\right)}$ in Proposition \ref{ncopydistill} refers to positivity on states of Schmidt rank $2$, where the Schmidt rank is calculated w.r.t. the splitting $\mathcal{H}_N^{\otimes n}=\mathcal{H}^{\otimes n}_K\otimes\mathcal{H}^{\otimes n}_M$ of the multipartite space. This is important to notice because many different splittings of $\mathcal{H}_N^{\otimes n}$ into a tensor product of two factors are possible. Evidently, Proposition~\ref{ncopydistill} is nothing but Proposition~\ref{1copydistill} applied to $\rho^{\otimes n}$ in place of $\rho$. This is easy to understand because the tensor product $\rho^{\otimes n}$ represents a number $n$ of identical, independent copies of the state $\rho$, e.g. coming from a source that produces $\rho$.

It is natural to mention here a fundamental question concerning distillability of quantum states. Using the language of numerical ranges, we can formulate the problem in the following way: 
\vskip 2 mm
\noindent Given a density matrix $\rho$ on a bi-partite Hilbert space $\mathcal{H}_N^{\otimes n}=\mathcal{H}^{\otimes n}_K\otimes\mathcal{H}^{\otimes n}_M$ s.t. $\Lambda\left(\left(\1\otimes T\right)\rho\right)\not\in\left[0;+\infty\right)$, can we infer that $\Lambda^{\left(2\right)}\left(\left(\1\otimes T\right)\rho^{\otimes n}\right)\not\in\left[0;+\infty\right)$ for some positive integer $n$? 
\vskip 2 mm
\noindent In other words, is a bi-partite state $\rho$ with a negative partial transpose always distillable, possibly using a huge number $n$ of copies of $\rho$? This question has not yet been answered, despite a considerable effort and some partial results (cf. e.g. \cite{Pankowski}).

\subsection{Minimum output entropy and product numerical range}
Consider a completely positive map $\Phi$
acting on the set $\Omega_N$
of normalized quantum states of dimension $N$.
Minimum output entropy (see \eg\ \cite[Chapter 7]{petz}) is defined as
\begin{equation}
\MOE{\Phi} = \min_{\rho} \left\{ \vE{\Phi(\rho)}\right\},
\end{equation}
with $\rho\in\Omega_N$. 
Since the von Neumann entropy is concave, 
the minimum is attained on the boundary and thus
\begin{equation}
\MOE{\Phi} = \min_{\proj{\psi}} \left\{ \vE{\Phi(\proj{\psi})}\right\},
\end{equation}
where $\proj{\psi}\in\Omega_N$ are pure states. Therefore 
the minimum output entropy
can be interpreted as a certain measure
of decoherence 
introduced by the channel.

In \cite{king01minimal,king02additivity} it was proven that  minimum output 
entropy (and thus Holevo capacity) is additive for unital channels. It is now 
known however that minimum output entropy is not additive in the general 
case~\cite{hastings09counterexample}.
%
%
Here we provide a characterization of the minimum output entropy for 
one-qubit channels using product numerical range of the dynamical matrix.

\begin{proposition}
\label{prop4}
Let $\Phi$ be a CP-TP map acting on $\States{2}$. Then
\begin{equation}
 \MOE{\Phi} = \lambda \log (\lambda) + (1-\lambda) \log (1-\lambda),
\end{equation}
where $\lambda$ is a minimal value of product numerical range for the dynamical matrix 
$D_{\Phi}$ 
\begin{equation}
\lambda = \lambda^{\otimes}_{\min} (D_{\Phi}).
\end{equation}
\end{proposition}

\proof
Let us define $f(x) = -x\log_2(x)-(1-x)\log_2(1-x)$, which is increasing for 
$x \in [0,\frac{1}{2}]$. 

Directly from the definition of minimum output entropy we can write
\begin{eqnarray}
\MOE{\Phi} = \min_{\ket{i}} S(\Phi(\ketbra{i}{i}))
= \min_{\ket{i}} f( \lambda_{\min}(\Phi(\ketbra{i}{i}))) 
= f\left(\min_{\ket{i}}\lambda_{\min}(\Phi(\ketbra{i}{i}))\right).
\end{eqnarray}
Now since $\bra{k} \Phi(\ketbra{i}{j})\ket{l} = \bra{k \otimes i} D_{\Phi}\ket{l \otimes j}$ 
(see \cite[Eqn. 11.25]{BZ06}) we can rewrite the above expression as
\begin{eqnarray}
\MOE{\Phi} = f\left(\min_{\ket{i},\ket{j}} \bra{j}\Phi(\ketbra{i}{i})\ket{j}\right) 
= f\left(\min_{\ket{i},\ket{j}} \bra{j \otimes i}D_{\Phi}\ket{j \otimes i}\right) 
= f\left(\lambda^{\otimes}_{\min} (D_{\Phi})\right).
\end{eqnarray}
\halmos

Using the above proposition, we can easily calculate minimal output entropy for 
channels listed below.

First we consider the amplitude damping, phase damping, phase flip, bit-flip and and 
bit-phase flip channels. In all those cases we can see from the Kraus form that the 
spectrum of the dynamical matrix has two zero eigenvalues. 
Then the plane spanned by the two eigenvectors corresponding to the 
zero eigenvalue contains at least one product state \cite{CMW08}.
Thus $\lambda_{\min}^{\otimes}(D_{\Phi}) 
= 0$ and the minimum output entropy for this channels is equal to zero.
Using Proposition \ref{prop4}
one can also easily calculate minimum output entropy 
for some other one qubit channels.
%
 Consider the  Werner-Holevo channel, 
described by the following dynamical matrix,
 \begin{equation}
 D_{\Phi_{\mathrm{HW}}} = \left(
 \begin{array}{cccc}
  \frac{p+1}{2} & 0 & 0 & 0 \\
  0 & \frac{1-p}{2} & p & 0 \\
  0 & p & \frac{1-p}{2} & 0 \\
  0 & 0 & 0 & \frac{p+1}{2}
 \end{array}
 \right),
 \end{equation}
 for $p \in [-1,1/3]$. 
  In this case $\lambda_{\min}^{\rm loc}(D_{\Phi_{\mathrm{HW}}}) 
 = \frac{1}{2}(1 - |p|)$ and thus
 \begin{eqnarray}
  \MOE{\Phi_{\mathrm{HW}}} &=&  - \frac{1}{2}(1 - |p|) \log_2 \frac{1}{2}(1 - |p|) 
   - \frac{1}{2}(1 + |p|) \log_2 \frac{1}{2}(1 + |p|)\\ 
 &=& - \frac{\log \left(\frac{1}{4}-\frac{p^2}{4}\right)+2 p \tanh ^{-1}(p)}{\log(4)}.
 \end{eqnarray}

In the case of higher dimensional quantum channels we can use properties
of product numerical range to check, whether for a given channel
its minimal output entropy is equal zero.

\begin{proposition}
For any completely positive, trace preserving (CP-TP) map we have
\begin{equation}
 \MOE{\Phi} = 0 \quad \text{ iff } \quad 1 \in \Lambda^{\otimes}(D_{\Phi}).
\end{equation}
\end{proposition}
\proof
Since $1 \in \Lambda^{\otimes}(D_{\Phi})$, there exists $\ket{i},\ket{j}$ such that 
\begin{equation}
 1 = \bra{i \otimes j} D_{\Phi} \ket{i \otimes j} = \bra{i}\Phi(\ketbra{j}{j})\ket{i}.
\end{equation}
Because $\Phi$ is CP-TP channel, we have $\tr \Phi(\ketbra{j}{j}) = 1$ and thus
\begin{equation}
\Phi(\ketbra{j}{j}) = \ketbra{i}{i}.
\end{equation}
The proposition follows.
\halmos

\subsection{Local discrimination of unitary operators}\label{sec:local-discr}

The problem of local distinguishability of multipartite quantum states was analyzed by Walgate \etal~\cite{WSHV00}.
Following their work, Duan \etal\ have shown \cite{DFY08}
that two unitary operations $U_1$ and $U_2$ are locally distinguishable
iff $0 \in \ProductNumRange{V}$ where $V=U_1^{\dagger}U_2$.
If this is the case, then there exists a product state $| \psi\rangle=|\phi_A,\phi_B\rangle$ such that
the states $U_1|\psi\rangle$ and $U_2|\psi\rangle$ are orthogonal
and thus distinguishable.

Our results on product numerical range allow us to solve the problem of local distinguishability 
for a wide class of unitary operators. If the operator $V=U_1^{\dagger}U_2$
has the tensor product structure, $V=V_1 \otimes V_2$,
the two unitaries $U_1,U_2$ are distinguishable iff the numerical range of any of the factors $V_1,V_2$
contains zero. This is the case when $0$ belongs to the convex hull of the spectrum of
the factor $V_1$ or of the factor $V_2$.

\begin{figure}[htp]
 \begin{center}
 \includegraphics[scale=0.6]{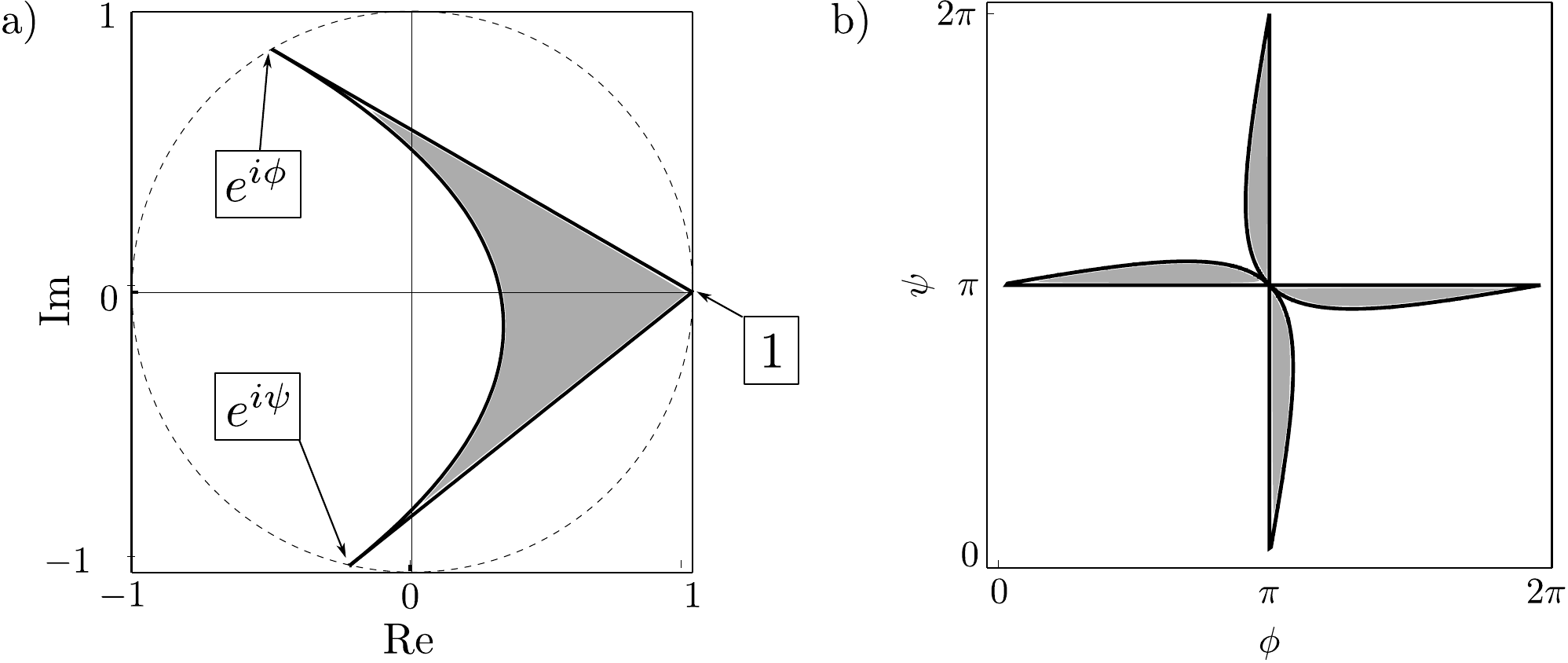}
\caption{a)\label{fig:localrange2b2} Product numerical range for the matrix 
\eqref{deffamily2b2} with $\phi= 2\pi / 3$ and $\psi= 10\pi / 7$, 
b)\label{fig:ufour}
 The region in the space of parameters $\left(\phi,\psi\right)$ corresponding 
to locally distinguishable 
pairs $\left(U_1,U_2\right)$ when $U_1^{\dagger}U_2=V\left(\phi,\psi\right)$.}
 \end{center}
\end{figure}

Let us now deal with a more general case of $V$ without the tensor product structure. 
Consider for instance a family of unitary matrices of order four, 
\begin{equation}
\label{deffamily2b2}
 V \left(\phi,\psi\right) = \left[\begin{array}{cccc}
	1&0&0&0\\
	0&e^{i\phi}&0&0\\	
	0&0&e^{i\psi}&0\\
	0&0&0&1
\end{array}\right],
\end{equation}
for $\phi,\psi\in\left[0,2\pi\right]$.

It is easy to show that the product numerical range of $V$ is a bounded region of $\mathbbm{C}$ whose border 
consists of the segments $\left[e^{i\phi},1\right]$, $\left[1,e^{i\psi}\right]$ and the 
line
\begin{equation}
 \gamma:\left[0,1\right]\ni t \ \mapsto \
t^2e^{i\phi}+\left(1-t\right)^2e^{i\psi}+2t\left(1-t\right)\in\mathbbm{C} \ .
\label{linegamma}
\end{equation}
For example, Fig. \ref{fig:localrange2b2}a shows the shape of the product
numerical range of $V(\frac{2\pi}{3}, \frac{10\pi}{7})$.

Using Eq. \eqref{linegamma}, it is not difficult to check for which values of the phases $\phi$ and $\psi$
the product numerical range of $V\left(\phi,\psi\right)$ contains $0$, so any $U_1$ and $U_2$ such that $U_1^{\dagger}U_2
=V\left(\phi,\psi\right)$ are locally distinguishable. 
Figure \ref{fig:ufour}b) shows, in grey, the set of parameters $\left(\phi,\psi\right)$ corresponding to such
 distinguishable pairs $\left(U_1,U_2\right)$. Explicitly, we have
\begin{multline}\label{distphipsi}
 0\in\LambdaRm^{\otimes}\left(V\left(\phi,\psi\right)\right)
\Leftrightarrow
 \Big\{|\sin\psi|\cos\phi+
 |\sin\phi|\cos\psi+2\sqrt{|\sin\phi\sin\psi|}\leqslant 0\\
 \land\sin\phi\sin\psi\leqslant 
 0\land\left(\phi,\psi\right)\not\in\left\{\left(0,0\right),\left(2\pi,2\pi\right)\right\}\Big\}.
\end{multline}

For any two unitary matrices $U_1$ and $U_2$
such that $V=U_1^{\dagger} U_2$ satisfies the above constraints,
it is possible to find a product state $|\chi, \xi \rangle $
with the property $\langle \chi, \xi | V |\chi, \xi \rangle =0 $.
A detailed construction of this state, presented in Appendix~\ref{sec:local-distinguish-vectors},
allows one to design the scheme of local discrimination 
between the unitary gates $U_1$ and $U_2$.

\subsection{Local fidelity and entanglement measures}
Several tasks of quantum information processing relay on the ability to approximate a given quantum state $\varrho_1$ 
by some other state $\varrho_2$. Alternatively, one attempts to distinguish $\varrho_1$ from $\varrho_2$.
To characterize both problems quantitatively one may use \emph{fidelity}, which can be interpreted as 
a `transition probability' in the space of quantum states \cite{Uh76},
\begin{equation}
\label{fidel1}
F(\varrho_1,\varrho_2) = [{\rm Tr}|\sqrt{\varrho_1}\sqrt{\varrho_2}|)]^2. 
\end{equation}
We are going to follow here the original definition by Jozsa \cite{Jo94},
but one has to be warned that some later articles use
the name `fidelity' for $\sqrt{F}$. 
If one of the states is pure, $\varrho_1=|\psi_1\rangle \langle \psi_1|$,
formula (\ref{fidel1}) simplifies and $F=\langle \psi_1 | \varrho_2 | \psi_1\rangle$.
Thus in this case fidelity has a simple interpretation of \emph{probability}
that the state $\varrho_2$ is projected onto a pure state $|\psi_1\rangle$.

 Consider two arbitrary mixed states $\varrho_1$ and $\varrho_2$
acting on a Hilbert space ${\cal H}_N$.
Although fidelity between these states is fixed and given by (\ref{fidel1}),
one may pose a question to what extent fidelity can grow if local unitary
operations are allowed. In other words, one asks about 
the fidelity between $\varrho_1$ and 
$U\varrho_2 U^{\dagger}$ maximized over all unitaries $U\in U(N)$. 
This problem was studied in \cite{MMPZ08}, where the 
 following bounds were established
\begin{equation}
 F ( p^{\uparrow}, q^{\downarrow}) \le 
 F (\varrho_1,U\varrho_2U^{\dagger}) \le
 F ( p^{\uparrow}, q^{\uparrow}) = F ( p^{\downarrow}, q^{\downarrow}).
\label{fidboth}
\end{equation}
The vectors $p$ and $q$ represent the spectra of $\varrho_1$ and $\varrho_2$,
while the up/down arrows
indicate that the eigenvalues are put in the non-decreasing (non-increasing, resp.) order.
Arguments of the fidelity in the above equation denote thus diagonal matrices
which represent classical states. 

In this section we analyze an analogous problem 
for multipartite systems: What maximum fidelity between
two given states of such a system can be achieved,
if arbitrary local unitary operations are allowed?
We provide a solution of this problem in the special case when
both quantum states are pure and derive bounds
for the local fidelity in the case where 
$\rho$ is a diagonal mixed state.

Let $|\phi\rangle$ be a vector and $\varrho$ an arbitrary 
mixed state, both on ${\cal H}_{AB}={\cal H}_{A} \otimes {\cal H}_{B}$.
For simplicity we will restrict our attention to the symmetric case
and assume that ${\rm dim}({\cal H}_{A})={\rm dim}({\cal H}_{B})=N$.

The fidelity of a mixed state with respect to a pure state is given by an expectation 
value, $F=\langle \phi | \varrho | \phi\rangle$. We are going to study the question
 to what extend this quantity can be increased by applying arbitrary
 local unitary operations $U_A\otimes U_B$. In other words, we look
 for the \emph{local fidelity} defined as the maximum

\begin{equation}
\label{locfid}
F^{\rm max}(\varrho,\phi)=\max_{U_A \otimes U_B} \langle \phi | 
(U_A \otimes U_B)^{\dagger} \varrho (U_A \otimes U_B) | \phi \rangle.
\end{equation}

It is instructive to relate this quantity to a generalized numerical radius
of an operator $X$, defined as the largest modulus of an element 
of its numerical range. Similarly for an operator $X$ acting on a
composed Hilbert space one defines product numerical radius
as the largest modulus of an element of $\ProductNumRange{X}$.
This notion can be further generalized, 
and for any operator $X$ and an auxiliary operator $C$ acting on the 
Hilbert space
 $\mathcal{H}_N = \mathcal{H}_K \otimes \mathcal{H}_M$,
one defines  the $C$--product numerical radius \cite{thomas08significance},
\begin{equation}
r_C^{\otimes}(X) = \max\{|z|:z=
 \tr (U_1\kron U_2)X(U_1\kron U_2)^{\dagger}C ,  
 \quad U_1\in U(K), \quad U_2\in U(M) \}, 
\end{equation}
and other  notions listed in  Tab. \ref{tab2}. 
The problem of finding the local fidelity is  then equivalent to
determining the $C$--product numerical radius 
of the operator $X= |\phi\rangle \langle \phi|$ for $C=\varrho$.

Let us first solve the problem in the special case where 
the analyzed state is pure, $\varrho=|\psi\rangle \langle \psi|$. 
It is then useful to represent both pure states using
their Schmidt decompositions (\ref{Aij}),
\begin{equation}
\label{Schmidt3}
|\phi\rangle =\sum_{i=1}^N \sqrt{\lambda_i} |i\rangle \otimes |i\rangle ,
{\quad}
|\psi\rangle =\sum_{j=1}^N \sqrt{\mu_j} |j\rangle \otimes |j\rangle .
\end{equation}
The vector $\lambda$ of Schmidt coefficients set in a decreasing (increasing) order 
will be denoted by $\lambda^{\downarrow}$ and $\lambda^{\uparrow}$, respectively.
This notation allows one to formulate the following lemma. 

\begin{lemma}
For arbitrary local unitary operation $U_{A} \otimes U_{B}$
and pure states $|\phi\rangle$, $|\psi\rangle$, one has
\begin{equation}
\label{bound7}
0 \leq |\langle \psi |U_{A} \otimes U_{B} | \phi \rangle|^{2} 
\leq F(\mu^{\downarrow},\lambda^{\downarrow})=
\left( \sum_{j=1}^N \sqrt{ \lambda_j^{\downarrow} \mu_j^{\downarrow}}\right)^2 .
\end{equation}
\end{lemma}

The lower bound is a trivial consequence of the definition of fidelity.
The upper bound follows from 
the theorem of Uhlmann which states that fidelity is given by the maximal overlap 
between purifications of both states,
and the bound in~\cite[Eq. (4.19)]{MMPZ08}.
This result follows also from the recent work of 
Schulte-Herbr{\"u}ggen \etal\ \cite[Prop. IV.1]{SHGDH08}.
 
If one of the states is separable, $|\phi\rangle=|\phi_A \rangle \otimes |\phi_B \rangle$, 
its Schmidt vector has only a single non-vanishing component, 
$\lambda^{\downarrow}=(1,0,\dots,0)$, so 
the overlap (\ref{bound7}) is bounded by the largest 
Schmidt coefficient $\mu_{\rm max}$ of the state $|\psi\rangle$.
This is a special case of the
{\it geometric measure of entanglement} of a multipartite state $|\psi\rangle$,
defined as the logarithm of the maximum projection on any product state \cite{WG03}, 
\begin{equation}
\label{geom}
E_g(|\psi\rangle) \ = - {\rm log} \Bigl( 
\max_{U_{\rm loc}}
|\langle \psi| U_1 \otimes U_2 \otimes \cdots \otimes U_m|0, \dots 0\rangle |^2 \Bigr) \ . 
\end{equation}
Here $|0, \dots 0\rangle$ represents an arbitrary product state, 
so transforming it by a local unitary matrix 
one explores the entire set of separable pure states of the $m$--partite system.
Observe that the argument of the logarithm in 
the above expression is just equal to the product numerical radius of the projector
onto the analyzed state, $X=| \psi\rangle \langle \psi |$.

In recent papers \cite{HMMOV09,WS09}
it was shown that the above maximization procedure
becomes simpler if the multipartite state $|\psi\rangle$ 
is symmetric with respect to permutations of the subsystems,
and all its coefficients in the product basis are non-negative.
Then the maximum in (\ref{geom}) is achieved for the tensor product
of a single unitary matrix, $U_{\rm loc}=U^{\otimes m}$, 
so the search for $E_g(\psi)$ can be reduced to 
the space of a smaller dimension.
It is then natural to ask whether this observation can be
generalized for the problem of determining the product
numerical radius of any multipartite Hermitian operator $X$,
provided that $X$ is symmetric with respect to permutations
and it satisfies suitable positivity conditions.
This problem was considered in a~very recent paper 
by H\"{u}bner \etal \ \cite{HKWG09}.

Thus the product numerical radius is useful in characterizing quantum entanglement
of a pure state of a multipartite system. Interestingly, the product $C$--numerical
radius of a Hermitian bi-concurrence matrix 
introduced by Badzi{\c a}g et al. \cite{BDHHH02}
can be applied to describe the degree of quantum entanglement
for any mixed state of a bipartite system.

Let us then return to the bipartite problem and 
discuss the case when one of the two states
in the expression (\ref{locfid}) for local fidelity is pure while the other is mixed.
Assume that the pure state $\ket{\phi}$
is given by its Schmidt decomposition (\ref{Schmidt3}),
while $\varrho$ is a diagonal mixed state, 
$\varrho = \sum_{i j = 1}^N p_{ij} |i\rangle \langle i| \otimes |j\rangle \langle j|.$
The maximal local fidelity between these states 
can be bounded by the following lemma, proved in Appendix \ref{sec:app-c}.

\begin{lemma}
 \label{lemma:lemmaC}
The maximal fidelity between a pure state $\psi$ and diagonal state $\rho$ is bounded from above,
\begin{equation}
\label{eqn:maxmax}
\max_{U, V \in U(N)} F\left((U \otimes V)\ketbra{\psi}{\psi} (U \otimes V)^{\dagger}, \varrho\right)
\leq \max \sum_{ij=1}^N p_{ij} B_{ij},
\end{equation}
where the maximum 
on the right-hand side is taken over all collections of non-negative
real numbers $B_{ij}$ that satisfy the constraints, 
for any $\{i_1, i_2, \dots, i_r\} \subset \{1,2, \dots , N\}$
\begin{equation}\label{constraint:sumI}
\sum_{j=1}^N \left[ B_{i_1 j} + B_{i_2 j} + \dots + B_{i_r j} \right]
\in \left[ \sum_{k=1}^{r} \lambda_{(k)} , \sum_{k=N-r+1}^{N} \lambda_{(k)} \right]
\end{equation}
and for any $\{j_1, j_2 , \dots , j_s \} \subset \{1,2, \dots , N\}$
\begin{equation}\label{constraint:sumJ}
\sum_{i=1}^N \left[ B_{i j_1} + B_{i j_2} + \dots + B_{i j_s} \right]
\in \left[ \sum_{k=1}^{s} \lambda_{(k)} , \sum_{k=N-s+1}^{N} \lambda_{(k)} \right],
\end{equation}
where $\lambda_{(1)}, \lambda_{(2)}, \dots , \lambda_{(N)}$ are Schmidt coefficients of $\psi$ in ascending order.
\end{lemma}

The maximum on the right-hand side in Lemma~\ref{lemma:lemmaC}
is attained at the edges of the polygon defined
by the constraints (\ref{constraint:sumI}) and (\ref{constraint:sumJ}).
The bounds obtained in this way can be easily computed numerically using the simplex algorithm.

\subsection{Local dark spaces and error correction codes}
Consider a quantum operation $\Phi$ acting in the space of mixed quantum states 
of size $N$, which can be represented in the Kraus form
\begin{equation}
\label{kraus1}
\rho' = \Phi(\rho) = 
\sum_{i=1}^M Y_i \rho Y_i^{\dagger}.
\end{equation}
To assure that the trace is preserved by the operation,
the set of $M$ Kraus operators has to satisfy 
an identity resolution,
$\sum_{i=1}^M Y_i^{\dagger} Y_i = {\mathbbm 1}$.

Consider a $l$-dimensional subspace $P_l = \sum_{i=1}^l \ket{i}\bra{i}$
embedded in ${\cal H}_N$.
If it satisfies the set of $M$ conditions
\begin{equation}
\label{compr1}
 P_l X_m P_l = \lambda_m P_l\; , {\rm \quad for \quad} m=1,\ldots,M
\end{equation}
where $X_m=Y_m^{\dagger} Y_m$ and $\lambda_m \in \Cplx$ 
no information goes outside of this subspace \cite{MK06},
so $P_l$ is called a~\emph{dark subspace} \cite{MMZ09}.

If a subspace $P_k$ fulfils even stronger conditions of the type (\ref{compr1}),
\begin{equation}
\label{compr2}
 P_l Y_i^{\dagger}Y_j P_l = \lambda_{ij} P_l\; , {\rm \quad for \quad} i,j=1,\ldots,M,
\end{equation}
then quantum information stored in the system can be recovered, 
so the subspace $P_l$ provides an error correction code \cite{BDSW96a,KL97}.
Note that $P_l$ has to simultaneously satisfy all the $M^2$
equations (\ref{compr2}). The complex numbers $\lambda_{ij}$ 
corresponding to different $X_{ij}$'s may be different.

From an algebraic perspective condition (\ref{compr1}) 
implies that $\lambda_m$ belongs to the numerical range of order $l$
of the operator $X_m$ \cite{CKZ06}.
In full analogy to the product numerical range, one may 
introduce the concept of  {\it product numerical range of higher rank}
as defined in  Tab.  II. 
This notion can be used to identify dark spaces or error correction codes
with a~local structure \cite{DH09}.
The distinguished subspace $P_l^{\otimes}$,
which solves the set of equations (\ref{kraus1}),
can be chosen to be in the product form,
$P_l^{\otimes} = \sum_{i=1}^l \ket{i\kron i}\bra{i \kron i}$.

\section{Concluding remarks}
\label{sec:concl}

In this work we investigated basic properties of numerical range
of an operator restricted to some class of quantum states.
In particular, we analyzed the case of operators 
acting on a Hilbert space with a tensor product structure,
often used to describe composed quantum systems.
In this case one defines the product numerical range of an operator.
We reviewed basic properties of this notion
and presented some examples of operators
for which product numerical range can be found analytically.




To tackle the problem in a general case, however, we had to rely
on numerical computations. In particular, we investigated an ensemble of 
$N=4$ random density matrices distributed according to the Hilbert-Schmidt
measure and compared the probability distributions of both edges of the product range
with probability distributions for individual eigenvalues.

In the case of a non-Hermitian operator its product numerical range forms a connected set
in the complex plane. In general this set is not convex.
The product numerical range of  an operator acting on a two-fold tensor product
is simply connected. However, this property does not hold 
for operators acting on a space with a larger number of subsystems.
For any operator with a tensor product structure
its product range is equal to the Minkowski product of numerical ranges
of all factors. The theory of the Minkowski product of various sets in the
complex plane, recently developed by Farouki \etal\ \cite{FMR01},
can thus be directly applied to characterize the product numerical range
of operators of the tensor product form. In this way we managed to establish
product numerical range of a unitary product matrix $U^{\otimes n}$.




\begin{table}
\begin{tabular}{|c|c|}
\hline
\hline
Standard definitions  & Product definitions 
\\ 
(for simple systems) & (for multipartite systems) \\
$X:\mathcal{H}_N \to \mathcal{H}_N $ & 
$X:\mathcal{H}_{n_1} \otimes \dots \otimes \mathcal{H}_{n_m}
\to 
\mathcal{H}_{n_1} \otimes \dots \otimes \mathcal{H}_{n_m}$
\\ \hline \hline
numerical range & product numerical range \\
$\NumRange{X}=\left\{\bra{\varphi}X\ket{\varphi}: \ket{\varphi}\in\mathcal{H} \right\}$ 
&
$\ProductNumRange{X}=\left\{\bra{\psi_1\otimes\ldots\otimes\psi_m}X
\ket{\psi_1\otimes\ldots\otimes\psi_m} :
\ket{\psi_i}\in\mathcal{H}_{n_i}\right\}$ 
\\ \hline 
numerical radius &  product numerical radius \\
$r(X) = \max\{|z|:z\in\NumRange{X}\}$ 
&
$r^{\otimes}(X) = \max\{|z|:z\in\ProductNumRange{X}\}$ 
\\ \hline
$C$-numerical range& product $C$-numerical range\\
$\CNumRange{X}=\{\lambda: \lambda=\tr UXU^{\dagger}C \}$ 
&  
$\LambdaRm_C^\otimes(X)=
\left\{ \tr (U_1\kron\ldots\kron U_k)X(U_1\kron\ldots\kron U_k)^{\dagger}C \right\} $ 
\\
where $U\in U(N)$. 
& 
where $U_i\in U(n_i)$.
\\ \hline
$C$-numerical radius & product $C$-numerical radius \\
$r_C(X) = \max\{|z|:z\in\CNumRange{X}\}$  
& 
$r_C^{\otimes}(X) = \max\{|z|:z\in\LambdaRm_C^\otimes(X)\}$ 
\\ \hline
higher rank numerical range & higher rank product numerical range \\
$\LambdaRm_l(X) = \{ \lambda: P_l X P_l = \lambda P_l \} $ 
& $\LambdaRm_l^\otimes(X) = \{ \lambda: P_l^\otimes X P_l^\otimes = \lambda P_l^\otimes \}$ 
\\
$P_l = \sum_{i=1}^l \ket{i}\bra{i}$ & $P_l^\otimes = \sum_{i=1}^l \ket{\underbrace{i\kron\ldots\kron i}_m}\bra{\underbrace{i\kron\ldots\kron i}_m}$
\\ \hline \hline
\end{tabular}
\caption{
Standard algebraic definition of the numerical range 
and related concepts compared with 
their \emph{product analogues}.
The definitions on the left concern an operator $X$ acting on Hilbert space 
$\mathcal{H}_N$, while their product analogues are defined 
for operators acting on a tensor product Hilbert 
space $\mathcal{H}_{n_1} \otimes \dots \otimes\mathcal{H}_{n_m}$.
Here $\ket{\varphi}$ denotes an 
arbitrary state of $\mathcal{H}_N$, while $\ket{\psi_1\otimes \dots\otimes \psi_m}
=\ket{\psi_1} \otimes \dots \otimes \ket{\psi_m}\in
 \mathcal{H}_1 \otimes \dots \mathcal{H}_m$ represents an arbitrary 
product state of the composite, $m$-particle  system.}
\label{tab2}
\end{table}


Numerical range can also be generalized by taking other restrictions
on the set of quantum states. Although we studied here 
the case of numerical range restricted to  separable 
and $k$--entangled states, one may 
 also use other restricted sets of quantum states
 or combine these conditions, analyzing for instance the set of 
real product states. As the product states of the $K\times M$ system
can also be considered as coherent states with respect to the
composite group $SU(K) \otimes SU(M)$ \cite{MZ04},
an analogous relation holds for the corresponding numerical ranges.

Numerical range can also be generalized in other direction:
for each case of a restricted numerical range one can 
introduce concepts  and generalizations
known for the standard numerical range.
In Table \ref{tab2} we have collected standard definitions of numerical
range, numerical radius, $C$--numerical range and higher rank numerical range
\cite{CKZ06}, along with their counterparts defined for Hilbert space of the
form of an $m$--fold tensor product,
$\mathcal{H}_N
= \mathcal{H}_{n_1} \otimes \cdots \otimes \mathcal{H}_{n_m}$,
 with $N=n_1\dots n_m$. Note that C-numerical range, as well as product C-numerical
range, reduce to the numerical range (product numerical range, resp.) for
$C = \diag (\{1,0,\dots ,0\})$ and this case was already analyzed in
\cite{dirr08relative}.

Observe that the above concepts arise naturally
in a variety of problems  in quantum information theory.
For instance, being in a position to find the product numerical range 
of an arbitrary operator, one could advance fundamental
problems concerning the characterization of the set of positive maps 
or description of the set of entangled states
and finding the minimum output entropy of a one--qubit
quantum channel.
Therefore, improving techniques of finding restricted numerical ranges
would have direct implications for the theory of quantum information
and quantum control.
For example, in this work we have established 
the positivity of a certain family of one-qubit maps,
we solved the problem of local distinguishability 
between a class of two-qubit unitary gates
and analyzed the properties of local fidelity between quantum states.

In conclusion, we advocate further studies on
restricted numerical range and cognate concepts.
On one hand, the restricted numerical range is
an interesting subject for mathematical investigations.
On the other hand, it proves to be a~versatile algebraic tool,
useful in tackling various problems of quantum theory.

\begin{acknowledgements}
It is a pleasure to thank M.D.~Choi, P.~Horodecki, C.K.~Li and T.~Schulte-Herbr{\"u}ggen 
for fruitful discussions and
to G.~Dirr, J. Gruska, M. Huber, M.B. Ruskai, 
and M. Sot{\'a}kov{\'a} for helpful remarks.
We acknowledge the financial support 
by the Polish Ministry of Science and Higher Education under the grants number 
N519 012 31/1957 and DFG-SFB/38/2007,
and Project operated within the Foundation for Polish Science International Ph.D.
Projects Programme co-financed by the European Regional Development Fund covering,
under the agreement no. MPD/2009/6, the Jagiellonian University International
Ph.D. Studies in Physics of Complex Systems.
The numerical calculations presented in this work were performed on
the \texttt{Leming} server of The Institute of Theoretical and Applied
Informatics, Polish Academy of Sciences.
\end{acknowledgements}

\appendix
\section{Product vectors for local discrimination between $U_1$ and $U_2$}
\label{sec:local-distinguish-vectors}

The discussion in Section \ref{sec:local-discr} left aside the
question of precisely how the unitaries $U_1$, $U_2$ fulfilling $\ProductNumRange{U_1^{\dagger}U_2}=0$ can be distinguished. 
To accomplish this task in practice, one needs to find a product vector $|\chi, \xi \rangle$ such 
that $\langle\chi , \xi|U_1^{\dagger}U_2|\chi, \xi\rangle=0$. There exists in general an infinite number of such vectors.
In the case $U_1^{\dagger}U_2=V\left(\phi,\psi\right)$ analyzed in Section \ref{sec:local-discr}, it is not difficult to find all of them.
Recall that $V\left(\phi,\psi\right)$ is of the diagonal form $\textnormal{diag}\left(1,e^{i\phi},e^{i\psi},1\right)$ with respect to the 
tensor product basis $\left\{|00\rangle,|01\rangle,|10\rangle,|11\rangle\right\}$ of $\mathcal{H}_2\otimes\mathcal{H}_2$. Let us write
$|\xi\rangle=\sqrt{t}\,e^{i\kappa_0}|0\rangle+
\sqrt{1-t}\,e^{i\kappa_1}|1\rangle$ and $|\chi\rangle=\sqrt{s}\,e^{i\delta_0}|0\rangle+
\sqrt{1-s}\,e^{i\delta_1}|1\rangle$ for $s,t\in\left[0,1\right]$ and $\kappa_0,\kappa_1,\delta_0,\delta_1$ arbitrary real numbers. 
Thus we assume that $|\chi\rangle$ and $|\xi\rangle$ are of unit norm, which is permissible.
It is now easy to see that
\begin{equation}
\label{eqn:contr-prod}
 \langle\chi , \xi|V\left(\phi,\psi\right)|\chi, \xi\rangle =
ts+\left(1-t\right)\left(1-s\right)+e^{i\phi}t\left(1-s\right)+e^{i\psi}\left(1-t\right)s,
\end{equation}
where $s,t\in\left[0,1\right]$.

Formula \eqref{eqn:contr-prod} gives us some idea of how the results presented in
Section \ref{sec:local-discr} were obtained. 
Note that the phases $\kappa_0,\kappa_1,\delta_0$ and $\delta_1$ are irrelevant to the value
 of $\langle\chi, \xi|V\left(\phi,\psi\right)|\chi, \xi \rangle$.
Thus any product vector that fulfils certain relations between the amplitudes $t$ and $s$ can be used for perfect discrimination between
 the two unitaries. Note that this is a general property whenever $U_1^{\dagger}U_2$ is diagonal with respect to some tensor product basis 
of $\mathcal{H}_K\otimes\mathcal{H}_M$ and $0\in \ProductNumRange{U_1^{\dagger}U_2}$.

In order to solve Eq. (\ref{eqn:contr-prod}) for $s$ and $t$, we first observe that 
\begin{equation}
\textnormal{Im}\left(\langle\chi ,\xi|V\left(\phi,\psi\right)|\chi, \xi\rangle \right)=0
\end{equation} 
reduces to $\sin\phi t\left(1-s\right)+\sin\psi s\left(1-t\right)=0$ or
\begin{equation}\label{eqn:sasafunctionoft}
 s=\frac{t\sin\phi}{t\left(\sin\phi+\sin\psi\right)-\sin\psi}.
\end{equation}
If we substitute this in Eq. \eqref{eqn:contr-prod}, we get the condition 
$\langle\chi, \xi| V\left(\phi,\psi\right)|\chi, \xi\rangle =0$ 
in the following form
\begin{equation}
\label{eqn:zerorealpart}
 t^2\sin\phi+\left(1-t\right)\left(t\sin\left(\phi-\psi\right)-\left(1-t\right)\sin\psi\right)=0 .
\end{equation}
We can solve \eqref{eqn:zerorealpart} for $t\in\left[0,1\right]$ under the assumption
that $0\in\ProductNumRange{V\left(\phi,\psi\right)}$ (cf. the conditions
on the right-hand side of \eqref{distphipsi}). The result is
\begin{equation}\label{eqn:formulafort}
 t=\frac{\sqrt{\sin\left(\phi-\psi\right)^2+4\sin\phi\sin\psi}+|\sin\left(\phi-\psi\right)|+2|\sin\psi|}
{2\left(|\sin\phi|+|\sin\left(\phi-\psi\right)|+|\sin\psi|\right)}.
\end{equation}
By symmetry we obtain an expression for $s$,
\begin{equation}\label{eqn:formulafors}
 s=\frac{\sqrt{\sin\left(\psi-\phi\right)^2+4\sin\psi\sin\phi}+|\sin\left(\psi-\phi\right)|+2|\sin\phi|}
{2\left(|\sin\psi|+|\sin\left(\psi-\phi\right)|+|\sin\phi|\right)}.
\end{equation}
Hence the product vector useful for perfect local discrimination between $U_1$ and $U_2$ can be any of the family
\begin{equation}\label{eqn:prodvector}
\left(\sqrt{t}\,e^{i\kappa_0}|0\rangle+
\sqrt{1-t}\,e^{i\kappa_1}|1\rangle\right)\otimes\left(\sqrt{s}\,e^{i\delta_0}|0\rangle+
\sqrt{1-s}\,e^{i\delta_1}|1\rangle\right) , 
\end{equation}
with $\kappa_0,\kappa_1,\delta_0,\delta_1\in\R$ and $s,t$ given by the formulas \eqref{eqn:formulafors} and \eqref{eqn:formulafort}, respectively. 
This only works when $U_1^{\dagger}U_2=V\left(\phi,\psi\right)$ and $0\in
\ProductNumRange{U_1^{\dagger}U_2}$.

\section{Proof of Lemma \ref{lemma:lemmaC}}
\label{sec:app-c}

Let us introduce matrix $A$ which depends on the vector
$\lambda$ and a local unitary matrix $U \otimes V$, 
with entries
\begin{equation}\label{notationaij}
A_{ij} = \left|\sum_{k=1}^N \sqrt{\lambda_k} \scalar{U\left(k\right)}{i} \scalar{V\left(k\right)}{j} \right|^2,
\end{equation}
where by $U\left(k\right)$ we mean $U\left(\ket{k}\right)$, so that $\scalar{U\left(k\right)}{i}$ corresponds to $\bra{k} U^{\dagger}\ket{i}$ in the usual physicists' notation. Similarly, $\scalar{k}{U\left(i\right)}=\bra{k} U\ket{i}$.

Using the notation of eq. \eqref{notationaij}, we arrive at a handy expression for the expectation value 
\begin{eqnarray}
 \langle \psi| (U \otimes V) ^{\dagger} \rho \; (U \otimes V) | \psi \rangle &=&
\sum_{ij=1}^N p_{ij} A_{ij} ,
\end{eqnarray}
which we wish to maximize over the set of local unitaries.
The first thing to notice is that $A_{ij}$ are non-negative real numbers and
\begin{equation}
\sum_{ij=1}^N A_{ij} = 1 ,
\end{equation}
thus the matrix $A$ treated as vector is an element of standard $( N^2 - 1)$-simplex.

Matrix $A$ defined above satisfies the following lemma.
\begin{lemma}
For any $\{i_1, i_2 , \dots , i_r \} \subset \{1,2, \dots , N\}$
\begin{equation}
\sum_{j=1}^N \left[ A_{i_1 j} + A_{i_2 j} + \dots + A_{i_r j} \right]
\in [\lambda_{(1)} + \lambda_{(2)} + \dots + \lambda_{(r)} , \lambda_{(N)} + \lambda_{(N-1)} + \dots + \lambda_{(N-r+1)}]
\end{equation}
and for any $\{j_1, j_2 , \dots , j_s \} \subset \{1,2, \dots , N\}$
\begin{equation}
\sum_{i=1}^N \left[ A_{i j_1} + A_{i j_2} + \dots + A_{i j_s} \right]
\in [\lambda_{(1)} + \lambda_{(2)} + \dots + \lambda_{(s)} , \lambda_{(N)} + \lambda_{(N-1)} + \dots + \lambda_{(N-s+1)}] ,
\end{equation}
where $\lambda_{(1)}, \lambda_{(2)}, \dots , \lambda_{(N)}$ are the Schmidt 
coefficients of $|\psi\rangle$ sorted ascendingly.
\end{lemma}
\proof
First we write
\begin{eqnarray}
\nonumber
&&\sum_{j=1}^N \left[ A_{i_1 j} + A_{i_2 j} + \dots + A_{i_r j} \right]=\\
\nonumber
&&=\sum_{j=1}^N 
\left[
	|\sum_{k=1}^N \sqrt{\lambda_k} \scalar{U k}{i_1} \scalar{V\left(k\right)}{j} |^2
+ \dots
+ |\sum_{k=1}^N \sqrt{\lambda_k} \scalar{U\left(k\right)}{i_{r}} \scalar{V\left(k\right)}{j} |^2 \right] \\
\nonumber
&&=
\sum_{j=1}^N \Big[ \sum_{k_1 l_1=1}^N \sqrt{\lambda_{k_1} \lambda_{l_1}} \scalar{U\left(k_1\right)}{i_1} \scalar{V\left(k_1\right)}{j} \scalar{i_1}{U\left(l_1\right)} \scalar{j}{V\left(l_1\right)}
+ \\ 
\nonumber
&&\dots +
\sum_{k_r l_r=1}^N \sqrt{\lambda_{k_r} \lambda_{l_r}}
 \scalar{U\left(k_r\right)}{i_r} \scalar{V\left(k_r\right)}{j} \scalar{i_r}{U\left(l_r\right)} \scalar{j}{V\left(l_r\right)} \Big]
\nonumber
\\
\nonumber
&&=
\sum_{k_1 l_1=1}^N \sqrt{\lambda_{k_1} \lambda_{l_1}} \scalar{U\left(k_1\right)}{i_1} 
 \scalar{i_1}{U\left(l_1\right)} \sum_{j=1}^N \scalar{V\left(k_1\right)}{j} \scalar{j}{V\left(l_1\right)}+ \\
&&\dots +
\sum_{k_r l_r=1}^N \sqrt{\lambda_{k_r} \lambda_{l_r}} 
\scalar{U\left(k_r\right)}{i_r} \scalar{i_r}{U\left(l_r\right)} \sum_{j=1}^N \scalar{V\left(k_r\right)}{j} \scalar{j}{V\left(l_r\right)} .
\end{eqnarray}
Since $\ket{j}$ form a basis, we have the identity $\sum\ketbra{j}{j} = \1$. Thus
\begin{eqnarray}
&&\sum_{j=1}^N \left[A_{i_1 j} + A_{i_2 j} + \dots + A_{i_r j} \right]= \nonumber
\\
&&=
\nonumber
\sum_{k_1 l_1=1}^N \sqrt{\lambda_{k_1} \lambda_{l_1}} \scalar{U\left(k_1\right)}{i_1} \scalar{i_1}{U\left(l_1\right)} \scalar{V\left(k_1\right)}{V\left(l_1\right)}+\\
&& \dots +
\sum_{k_r l_r=1}^N \sqrt{\lambda_{k_r} \lambda_{l_r}} \scalar{U\left(k_r\right)}{i_r}
 \scalar{i_r}{U\left(l_r\right)} \scalar{V\left(k_r\right)}{V\left(l_r\right)} \nonumber
\\
&&=
\sum_{k_1=1}^N \lambda_{k_1} |\scalar{U\left(k_1\right)}{i_1}|^2
+ \dots +
\sum_{k_r=1}^N \lambda_{k_r} |\scalar{U\left(k_r\right)}{i_r}|^2 .
\end{eqnarray}
This fact, combined with Corollary 4.3.18 of Horn and Johnson \cite{HJ1}, proves 
the lemma.
\halmos


\bibliographystyle{unsrt}
\bibliography{restricted-num-range}
\end{document}